\documentclass{aastex62}
\usepackage{amsmath,amssymb}
\usepackage{graphicx}
\usepackage{float}
\usepackage{mathtools}
\renewcommand{\b}[1]{\boldsymbol{#1}}  
\newcommand{\h}[1]{\boldsymbol{\hat{#1}}}  

\def\deg{\ifmmode^\circ\else$^\circ$\fi}
\def\arcsec{\ifmmode^{\prime\prime}\else$^{\prime\prime}$\fi}
\def\arcmin{\ifmmode^{\prime}\else$^{\prime}$\fi}

\def\earth{\ifmmode_{\mathord\oplus}\else$_{\mathord\oplus}$\fi}
\def\PUMA{{\tt PUMA}}
\def\PUMALINK{{\tt PUMALINK}}

\graphicspath{{./}{figures/}}

\received{Aug 13, 2023}
\revised{Sep 18, 2023}
\submitjournal{PASP}

\shorttitle{PUMA}
\shortauthors{Tonry}

\begin{document}

\title{Linking Sky-plane Observations of Moving Objects}

\correspondingauthor{John Tonry}
\email{tonry@hawaii.edu}

\author{J. L. Tonry}
\affiliation{Institute for Astronomy, University of Hawaii, 2680 Woodlawn Drive, Honolulu, HI 96822}

\begin{abstract}
The Asteroid Terrestrial-impact Last Alert System (ATLAS) observes the
visible sky every night in search of dangerous asteroids.  With four
(soon five) sites ATLAS is facing new challenges for scheduling
observations and linking detections to identify moving asteroids.
Flexibility in coping with diverse observation sites and times of detections
that can be linked is critical, as is optimization of observing time
for coverage versus depth.

We present new algorithms to fit orbits rapidly to sky-plane
observations, and to test and link sets of detections to find the ones
which belong to moving objects.  The \PUMA\ algorithm for fitting
orbits to angular positions on the sky executes in about a
millisecond, orders of magnitude faster than the methods currently in
use by the community, without sacrifice in accuracy.  The
\PUMA\ software should be generally useful to anyone who needs to test
many sets of detections for consistency with a real orbit.

The \PUMALINK\ algorithm to find linkages among sets of detections has
similarities to other approaches, notably HelioLinC, but it functions
well at asteroid ranges of a small fraction of an AU.  \PUMALINK\ is
fast enough to test 10 million possible tracklets against one another
in a half hour of computer time.  Candidate linkages are checked by the
\PUMA\ library to test that the detections correspond to a real
orbit, even at close range, and the false alarm rate is manageable.
Sky surveys that produce large numbers of detections from large
numbers of exposures may find the \PUMALINK\ software helpful.

We present the results of tests of \PUMALINK\ on three datasets which
illustrate \PUMALINK's effectiveness and economy: 2 weeks of all ATLAS
detections over the sky, 2 weeks of special ATLAS opposition
observations with long exposure time, and 2 weeks of simulated LSST
asteroid observations.  Detection probabilities of linkages must be
traded against false alarm rate, but a representative choice for
\PUMALINK\ might be 90\% detection probability for
real objects while keeping the false alarm rate below 10\%.
Although optimization of the tradeoffs between detection probability,
execution time, and false alarm rate is application specific
and beyond the scope of this paper, we provide guidance on methods
to distinguish false alarms from correct linkages of real objects.
\end{abstract}


\section{Introduction}

The Asteroid Terrestrial-impact Last Alert System (ATLAS) has been
funded by NASA since 2013 to warn of imminent, destructive asteroid
strikes on the Earth.   \cite{ATLAS18} described the ATLAS system
when it comprised two telescopes on two Hawaiian islands.
Since that time two new sites have been commissioned at Sutherland
Observatory in South Africa and at El Sauce in Chile, and a fifth ATLAS
unit is currently being brought online on Tenerife island at the Teide
Observatory \citep{2023arXiv230207954L}.

Asteroids smaller than $\sim$30~m will mostly explode harmlessly in
the atmosphere; larger impactors will generally cause significant
damage if they land near a populated area \citep{NRC2010}.  The civil
defense timescale required to act prior to a predicted impact is at
least several days.  These two facts determine what the ATLAS etendue
(product of sensitivity and area coverage rate) must be.  The
ATLAS sensitivity is chosen so that it can spot an approaching 30~m
asteroid about 3 days before impact\footnote{observing 90\deg\ from the
  Sun; warning time in other directions varies with the asteroid
  illumination}, and ATLAS now surveys the visible sky
four times every night (weather permitting).  These correspond to a
best conditions limiting magnitude of $m_{lim}\sim19.5$ per exposure
and about 105,000~deg$^2$/day (130,000~deg$^2$/day when ATLAS-5 is
fully operational).

Coordinating observations and linking detections from different sites
around the planet is challenging but offers unprecedented
opportunities.  ATLAS is in the process of updating its scheduler to
take advantage of this, and this paper introduceds a new, inter-day
linking algorithm so support this development.

There are a number of other ongoing sky surveys, notably
PanSTARRS-1 and PanSTARRS-2 \citep{2016arXiv161205560C} ($m_{lim}\sim22.5$ and
8,000~deg$^2$/day),
the Catalina Sky Survey ($m_{lim}\sim21.3$ and 3,600~deg$^2$/day)
\citep{crts},
the Zwicky Transient Facility ($m_{lim}\sim20.5$ and
40,000~deg$^2$/day) \citep{2019PASP..131a8002B, 2019PASP..131g8001G},
and ASASSN ($m_{lim}\sim18$ and 200,000~deg$^2$/day)
\citep{2017PASP..129j4502K}.
The LAST survey \citep{2023arXiv230404796O} is just getting started
but its wide field mode should produce
$m_{lim}\sim 21$ and 18,000~deg$^2$/day.
In a few years the Rubin Telescope will start its LSST survey
($m_{lim}\sim23.5$, the single exposure 5-sigma point source depth
averaged over all six filters weighted by the number of visits per
lunation) and 8,000~deg$^2$/day) \citep{2019ApJ...873..111I}.  Finding
dangerous asteroids is a significant mission for all of these surveys.

After data collection, surveys calibrate the astrometry and photometry
of each image, and then detect as many objects as possible, tabulating
their position, brightness, as well as attributes such as size and
shape.  Difference imaging, i.e. matching and subtracting a static sky
image taken some time previously, is an effective technique to support
the detection of transient, variable, or moving objects.
Implementations include \cite{1998ApJ...503..325A}, \cite{hotpants},
and \cite{zackaysub}.  This is helpful for suppressing detection of
static objects, but equally important is the reduction of the
influence of bright static objects on the astrometry and photometry of
non-static objects.  No matter the survey, the majority of detections
are very faint and appear at the signal-to-noise ratio (SNR) limit,
loosely defined as the point where the probability of detection is
distinctly less than unity, and the false alarm rate (random and
systematic) becomes significant.

When a survey collects multiple exposures, the probability of
detecting an object on one or more exposures is improved, and the
probability that false alarms, consistent with a single object, appear
on two or more exposures is reduced.  Deciding whether detections from
multiple exposures belong to a single object is the ``linking
problem''.  The uncertainty in the location of a point-like object is
approximately its full width half maximum (FWHM) extent divided by the
detection SNR, and for a stationary object the probability of a false
linkage between two observations is this solid angle times the angular
density of spurious detections.  Linking moving objects is more
difficult because the solid angle available for linking detections is
stretched by the angular velocity of the object times the interval
between exposures.  If the direction and speed of the object is
unknown this solid angle can be substantial.

There is an upper limit to an object's angular velocity that must be
considered.  If an object moves fast enough its appearance on an
exposure will be detectably trailed (roughly one FWHM per exposure
time at the SNR limit).  This rate is of order $\omega\sim1$~deg/day
(2.5~arcsec/min) for seeing limited surveys such as Pan-STARRS or
LSST, or $\omega\sim4$~deg/day for a pixel limited survey such as
ATLAS.  As a linking strategy, objects that are detectably trailed
should be sorted into a special subgroup for intercomparison, greatly
reducing the false alarm density.

False alarm detections arise from many sources: imperfect
subtractions, detector artifacts, cosmic rays, optical ghosts and
glints, satellite trails, etc, as well as detections of real objects
that happen to be coincident with a different object.  For example, a
typical detection density in an ATLAS difference image might be
200~deg$^{-2}$, of which only 10\% might be real objects.  It is
useful to apply a classification filter on detections before moving on
to linkage: most false detections are visibly false to a human
being, so machine learning approaches to classification can be
extremely effective \citep{2021PASP..133c4501C}.

The density of false alarms caused by real objects is highly variable.
For example \cite{2018AJ....156..241H} found the density of suspiciously
variable objects to be $n\sim$~20--1000~deg$^{-2}$, depending on galactic
latitude, to a limiting magnitude of about 18 away from the galactic plane
and 17 near the plane.

The physical density of known main belt asteroids in the ecliptic
plane (2.5~AU${<}r{<}$3.0~AU and $-0.2$~AU${<}z{<}$+0.2~AU) found in the {\tt
  astorb.dat} database\footnote{\tt asteroid.lowell.edu/main/astorb}
can be roughly described as
\begin{equation}
  n({<}D) \sim300,000\,\hbox{AU}^{-3}\; \left(D\over1\hbox{km}\right)^{-2}
\label{eq:nbelt}
\end{equation}
($D$ is the asteroid diameter), which integrates to an angular density
of about 150~deg$^{-2}$ for 1~km asteroids at a limiting magnitude of
21.5--23.5 (depending on viewing phase angle), or $\sim$4~deg$^{-2}$ 
for $\sim$6~km asteroids visible to ATLAS at opposition.  For evaluating
moving object densities at higher ecliptic latitude and angular
velocity, the density of asteroids near the Earth from {\tt astorb} is
approximately
\begin{equation}
  n({<}D) \sim150\,\hbox{AU}^{-3}\; \left(D\over1\hbox{km}\right)^{-2}
  \exp(-{|z|\over 0.14\hbox{AU}}).
\label{eq:neo}
\end{equation}
(Note that this is the instantaneous density; the count of all near
Earth objects (NEO) whose orbits bring them within 0.3~AU of the
Earth's orbit, for example, is about 6 times greater.)

It is more difficult to link moving objects at a finite distance with
non-contemporaneous observations because the observations need to
constrain an angular velocity as well as a position, and because the
object's unknown velocity allows confusion by false alarms from a
larger solid angle.

Any 3 coordinates on the sky supply the 6 constraints necessary to
formally solve for a three dimensional (3D) orbit that links them.
Orbits through unrelated detections can mostly be rejected because of
wildly unphysical velocities for example, or a sub-Earth surface
perigee between the three points, but there is no way to assess
goodness of fit for an orbit through three points and orbits which
differ in radial distance and velocity are highly degenerate.

If a survey covers a solid angle $\Omega$ in a night with a detection
angular density $n$, the number of pairs to be considered up to an
angular velocity $\omega$ over an inter-exposure time $\delta t$ is
$N_{pair}\sim\Omega\,\pi\,(\omega\delta t)^2\,n^2$, which can be
considerably greater than than 1 million for a typical survey.

Propagating each pair to the time of a third detection and requiring a
new detection at the extrapolated location multiplies the numbers by a
factor of $\pi\,\delta\theta^2\,n$, where $\delta\theta$ is the
typical astrometric uncertainty that grows over the interval from the
pair to the third (with allowance for all reasonable orbits).  This
factor grows quadratically with time and is typically very small for
time intervals less than an hour and very large for time intervals
greater than a day.

This factor is the crux of strategies to link detections of moving
objects.  On the one hand, collecting a third point quickly after a
pair lowers the false alarm rate, and collecting a fourth lowers the
false alarm rate again and permits a statistical evaluation of whether
the best-fit orbit is consistent with the observations.  Bright or
trailed detections may have a low enough angular density that only
3 detections are required.

On the other hand, relaxing the requirement for 4 quickly spaced
observations permits more sky to be covered, and requiring 4
detections can be costly for detection probability near the SNR limit.
For example, if there are 4 observations on which an object might be
detected with a probability of 0.8, the probability that it is in fact
detected 4 times is only 0.4.  The probability of obtaining at least 4
detections can be increased to 0.8 by taking 5 observations, but at
the cost of decreasing the time available to cover more sky by 20\%.

The Moving Object Processing System \citep{MOPS}, is used very
successfully by most asteroid surveys, including Pan-STARRS, the
Catalina Sky Survey, and ATLAS.  In its original design, when
Pan-STARRS was slated to comprise 4 telescopes on Maunakea, MOPS
expected to process pairs of detections from at least three nights
from a lunation.  (Naively this would consume $\mathcal{O}(N^3)$
computation time, where $N$ is the number of pairs, but this was
deemed to be computationally tractable.)

When Pan-STARRS became operational with only 1 telescope, the
combination of losses from fill factor, weather, and schedule made
this ``three tracklet'' approach unproductive, and Pan-STARRS switched
to an ``extrapolated arc'' approach, concentrating four exposures
into only an hour.  The ATLAS version of MOPS is also mindful of the
detection density and can officially declare a linkage for only 3
detections if they are bright or trailed enough.  However, the
majority of detections are faint and not detectably trailed, so a
full set of 4 detections is required.

A nice strategy for linking detections across a substantial time
interval arises when they are grouped as pairs of ``tracklets'' (a
tracklet is two or more linked detections), and explicitly using the
angular velocity of the tracklet as well as its location on the sky.
\cite{2000AJ....120.3323B} used this method to extend short arc
detections of Kuiper Belt objects (KBO).  The
\cite{2000AJ....120.3323B} approach was designed to guide new
observations to recover outer solar system objects with uncertain
orbits, and it exploited pairs of detections to provide a location and
a velocity.  Their motivation was linking very slow moving objects in
order to guide future observations with a small field of view.

More recently and generally \cite{2018AJ....156..135H} describe an
algorithm called ``HelioLinC'' to link pairs of detections over time
intervals of many days.  HelioLinC goes a step farther than
\cite{2000AJ....120.3323B} and links tracklets that may be moving
rapidly and not on a great circle.

HelioLinC's approach is to transform the detections on the sky to a 3D
location in the solar system, using the Sun as the origin, so that
object's trajectories lie on a great circle (neglecting Earth and Moon
gravity), with angular velocity consistent with a Keplerian orbit.
HelioLinC must assume a distance and radial velocity to perform this
transformation, but the subsequent linking problem becomes very simple
because each tracklet can be advanced along its orbit to a reference
time, and comparison of all tracklets at the reference time lends
itself to sorting and $N\log N$ execution time.

This comparison can be quick enough that it is possible to explore a
number of different assumptions for distance and radial velocity.
Because the tracklets can be actually propagated along an orbit, there
is no restriction that the observations of different tracklets be
particularly close in time --- matching two tracklets from anywhere
during a lunation is easy.  \cite{2018AJ....156..135H} demonstrated
the HelioLinC performance on objects ranging from the outer solar
system down to near Earth objects that are fairly close (closer than
1~AU).

Yet another approach by \cite{2021AJ....162..143M} called ``THOR''
links detections without requiring any organization into tracklets,
instead computing proximity relative to a set of presupposed trial
orbits.  THOR is especially useful for Main Belt asteroids where
Jupiter's gravity may be important because the distance makes the
angular variation among a modest number of trial orbits relatively
small.  It is considerably more difficult and resource consuming for
THOR to include enough NEO orbits to link objects that are close
to the Earth.  This is a disadvantage for the ATLAS mission of
finding and linking approaching impactors.

The unique advantage of THOR is the ability to link single detections,
but the price is that THOR is resource intensive.  The experiment
described by \cite{2021AJ....162..143M} linking a lunation of ZTF data
consumed 10M core-seconds, whereas the \PUMALINK\ algorithm described
here would complete in about 10k core-seconds, 3 orders of magnitude
faster.  However, without the organization of ZTF observations into
pairs separated by no more than a few hours, \PUMALINK\ would not work
at all.

Each of these algorithms, MOPS, HelioLinC, and THOR has
regimes of utility.  In its current incarnation MOPS requires 4
detections over a short period of time, but MOPS is highly efficient
and functions well for objects even closer than the Moon.  HelioLinC
requires only pairs of detections, but the tracklets can come from
different observatories and from times separated by days or weeks.
THOR does not even require observations organized into tracklets, but
is quite resource intensive.  Both HelioLinC and THOR make
approximations that hamper their effectiveness for asteroids which are
close to the Earth and feel the gravity of of Earth and Moon.

Neither MOPS nor HelioLinC is ideal for the ATLAS mission.  The false
alarm rate for faint pairs or triples inhibits ATLAS from submitting
them to the Minor Planet Center (MPC) for eventual linkage with other,
dubious pairs or triples, even though the combination may be
completely beyond question.  ATLAS also needs to recognize and link
very nearby asteroids, even closer than 0.04~AU (one week before
impact), and HelioLinC does not perform well in that regime.  In
addition, collection of 4 observations every night to satisfy MOPS is
costly.  If ATLAS can provide 3 days warning of a 30~m asteroid on a 1
day cadence by using MOPS and 4 detections per night, would it be not
be better to quadruple the exposure time but only take two
observations every other night?  The 2 day cadence can cost 1 day of
lost time, but the quadrupled exposure time means that the asteroid
will first be detected when it is 6 days distant, and its SNR will be
doubled when it is 3 days distant.  All that is required for this
improvement is a multi-day linkage scheme that is reliable at
distances less than 0.04~AU.

For many years ATLAS has used an algorithm called ``Position Using
Motion with Acceleration'' (\PUMA) for rapidly testing whether multiple
detections follow a consistent trajectory (i.e. orbit fitting).  This
code evolved from software that was originally developed to fit high
precision orbits to artificial satellites, so it does not use orbital
propagation approximations such as Keplerian, and it is
extremely efficient.  \PUMA\ has three notable features: it is very fast
(a handful of detections can be tested for a consistent orbital
trajectory in a millisecond), it takes full advantage of the
observational uncertainties, and it functions well at distances down
to the surface of the Earth.

We have recently extended the \PUMA\ approach to the linking problem,
calculating when pairs of pairs of detections form a consistent orbit.
The second section of this paper describes the \PUMA\ method for fitting
an orbit to a set of detections, the third section describes this new
\PUMALINK\ approach to linking pairs of detections, and the fourth section
details the results of a number of tests.

The \PUMA\ and \PUMALINK\ code may be found on
github\footnote{\tt https://zenodo.org/badge/latestdoi/674389255}$^,$\footnote{\tt https://github.com/atlas-ifa/puma/tree/main}.

\section{The PUMA library}

The \PUMA\ computation strategy is to determine a range from observer
to object for each observation using the Earth-Moon barycenter as the
origin.  Given the range of the object at each observation, it is
quick to determine an orbit that best threads through the (now) 3D
observed locations.  The displacement of the observer from the E-M
barycenter is known at all times, and the non-inertial motion of the
E-M barycenter is smooth enough to be described by a polynomial for a
substantial fraction of a year.

\subsection{Definitions}

\begin{figure}[h]
\begin{center}
\includegraphics[width=4in]{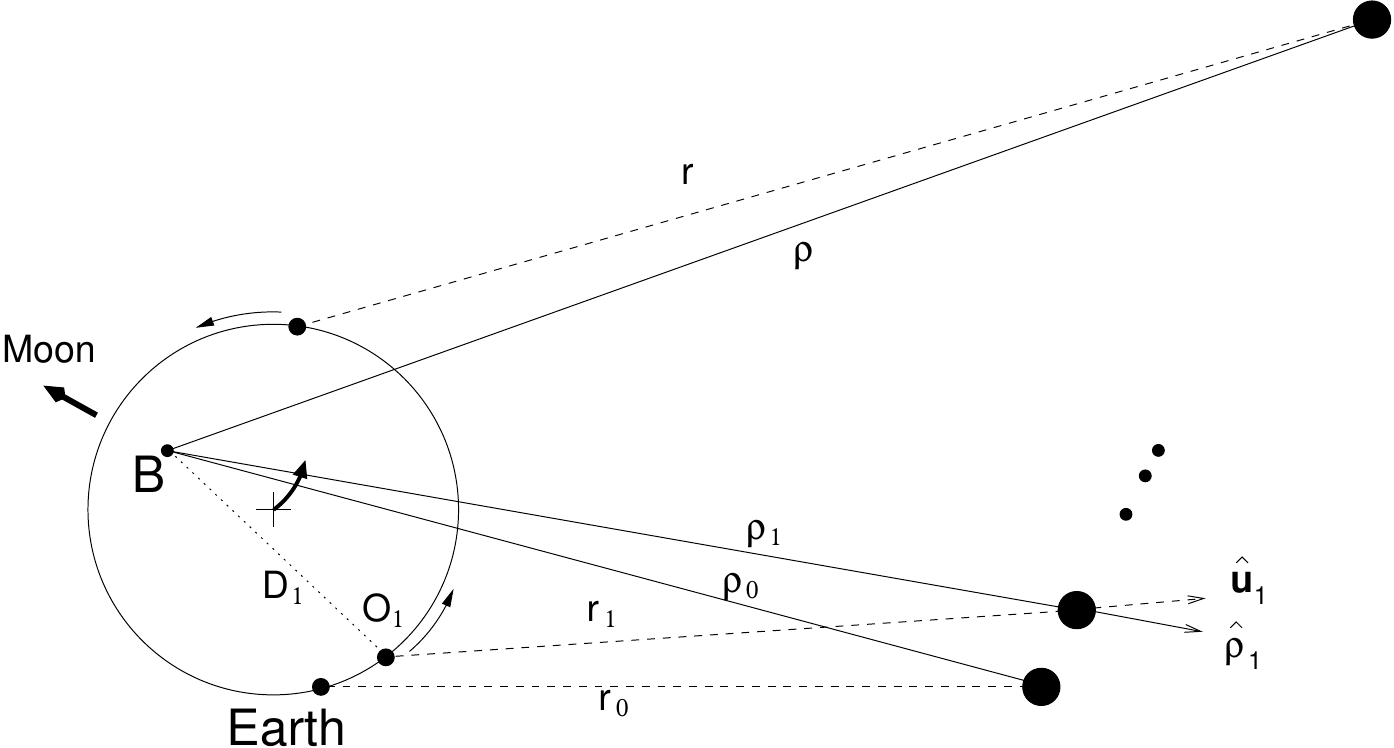}
\end{center}
\caption{The motion of an asteroid with respect to the Earth-Moon
  barycenter is viewed from an observatory on the rotating surface of
  the Earth which itself orbits the E-M barycenter.  The object is
  observed at ranges $r_i$ from the observatory locations $\b O_i^\prime$
  in directions $\b{\hat u}_i$ at times $t_i^\prime$.  The
  displacement from E-M barycenter to the observatory $\b D_i^\prime$
  is accurately known, and the distances and unit vector from
  barycenter to object are $\rho_i$ and $\b{\hat \rho}_i$.  (For clarity
  some variable's primes are omitted and the orbit of the non-inertial
  E-M barycenter around the Sun is not indicated.)}
\label{fig:pumavar}
\end{figure}

Starting with some definitions, suppose we have observations at
different times $t_i^\prime$ of an object that emitted light at
instants $t_i$.  Let the (known) observer locations in the solar
system at times $t_i^\prime$ be $\b O_i^\prime$, the (accurately
known, unaberrated) sight lines from observer at time $t_i^\prime$ to
object at time $t_i$ be $\h u_i$, and the (unknown) ranges between the
object at time of emission and the observer at time of observation be
$r_i$.  ($\h u_i$ and $r_i$ span events between emission and receipt
of light so we omit primes for their symbols.)  This is illustrated in
Figure~\ref{fig:pumavar}.

The Earth-Moon barycenter location in the solar system at time
$t_i^\prime$ is denoted as $\b B_i^\prime$, the displacement between
observation and barycenter is $\b D_i^\prime = \b O_i^\prime - \b
B_i^\prime$, and we describe the distance between E-M barycenter and
object as $\rho_i$.  (We ignore the few milliseconds of light travel time
between observer and barycenter, and we use {\it range} to describe
observer--object and {\it distance} for barycenter--object
displacements.)  Given a distance $\rho$ it is straightforward to
solve for the observer-object range using the law of cosines:
\begin{equation}
  r = \left[ \rho^2 - D_{\perp}^2 \right]^{1/2} - D_{\parallel},
\label{eq:rfromrho}
\end{equation}
where $D_{\perp}$ and $D_{\parallel}$ are the perpendicular and
parallel projections of $\b D^\prime$ relative to $\h u$.

The solar system locations of the object are therefore
$\b x_i = \b O_i^\prime + r_i \h u_i$, and the times are
related by $t_i = t_i^\prime - r_i/c$.  The unit vector from
E-M barycenter to object is
\begin{equation}
  \h \rho_i = {\b O_i^\prime + r_i\,\h u_i - \b B^\prime_i \over | \b O_i^\prime
    + r_i \,\h u_i - \b B^\prime_i |}.
\label{eq:bdot}
\end{equation}

As an example, given three observations of an object moving at
constant velocity in the solar system at a distance such that the
change in light travel time is small compared to the interval between
observations, it is simple to calculate the three ranges.  Define the
time ratio as $\lambda = (t_2-t_0)/(t_1-t_0)$ and the triple product
of the unit vectors as $T=\h u_0\times\h u_1\cdot\h u_2$.  Then
\begin{equation}
  \begin{pmatrix} r_0 \\ r_1 \\ r_1 \end{pmatrix} =
   {1\over \lambda\,(\lambda-1)\,T}
     \begin{pmatrix}
       -\lambda           \,\h u_1\times\h u_2 & \rightarrow \\
       (\lambda-1)       \,\h u_2\times\h u_0 & \rightarrow \\
      -\lambda(\lambda-1) \,\h u_0\times\h u_1 & \rightarrow
  \end{pmatrix}
  \begin{pmatrix}
    (\lambda-1)\,\b O_0^\prime - \lambda\,\b O_1^\prime +\b O_2^\prime \\
      \downarrow
  \end{pmatrix}.
  \label{eq:linsoln}
\end{equation}
The right arrows ($\rightarrow$) and down arrows ($\downarrow$)
indicate that the vector components are distributed into a row
or column respectively.  From this we immediately see that if the unit
vectors $\h u_0$, $\h u_1$, and $\h u_2$ are co-planar and $T=0$
(i.e. no relative acceleration between observer and object) the three
ranges $r_i$ are indeterminate.  A least squares solution for more
than three observations has the same property for co-planar $\h u_i$.

If $T\neq0$ it is possible to solve this equation, at least formally,
but the errors in the $\h u_i$ will be magnified by $1/T$, so we want
$T$ to be as big as possible, which requires some non-linearity
(acceleration) in the relative motion of observer and object.  Apart
from non-linear motion of the object, observations may happen from
different points on (or off) the Earth, the observer--object range has
a 24 hour sinusoidal component of amplitude $\sim$5,500~km from the
Earth's spin, the Earth's orbit around the Earth-Moon barycenter
creates a monthly sinusoidal component of order $\sim$4,700~km, and
the sagitta of the Earth's motion around the Sun over one day is
$\sim$5,700~km.  All of these contribute to non-linear motion of the
observer--object separation, but they take time to build up, which
makes the linking problem worse.

\subsection{PUMA}

The key to the \PUMA\ strategy is to determine the scalar distance
function $\rho(t)$ from E-M barycenter to object because it allows us
to place the angular observations $\b{\hat u}_i$ into 3D space where
it is easy to fit an orbit.  By design this function $\rho(t)$ can be
very well approximated by a polynomial in time for a fraction of a
year.  As we shall see, the choice of the Earth-Moon barycenter as the
origin instead of the Sun makes the result much less sensitive to the
distance and radial velocity of the object.  \PUMA\ calculates in
J2000 solar system ecliptic coordinates.

The \PUMA\ library proceeds iteratively to find the function
$\rho(t)$, based on assumed values for $\rho_0$ and $\dot \rho_0$ at a
reference time $t_0$.  The object's orbit may be very non-linear,
particularly when it passes close to the Earth, seen both in angular
observations that deviate from a great circle as well as distance
$\rho(t)$ that may so non-linear as to pass through a minimum.

The normal approach of a differential equation integrator is to start
from an initial condition and propagate in time by short, polynomial
steps that are brought into consistency by calculating the
acceleration at a set of test points.

By contrast \PUMA\ is exploiting a set of extremely accurate angular
positions at known times (to within slight perturbations from
differential light travel time), and it only calculates accelerations
at those times along the trajectory.  Rather than building a
piece-wise set of polynomials that can approximate an arbitrary
function, \PUMA\ uses an origin such that a single polynomial in
distance suffices.

\PUMA\ starts with a linear polynomial for the distances at the times
of observation, $\rho_i = \rho_0 + \dot \rho_0 (t_i-t_0)$, clipped at
$\rho_i>0.001\rho_0$.  Using these distances and therefore the 3D locations
${\b x}_i$ of the object in the solar system, the accelerations
$\b{\ddot x}_i$ of the object due to the gravity of the Sun, Earth,
and Moon are evaluated. (The location ${\b B}_i$, velocity $\b{\dot  B}_i$,
and acceleration $\b{\ddot B}_i$ of the E-M barycenter are
known.)

The second derivative of the displacement vector $\b\rho$ from E-M
barycenter to object location is just
$\b{\ddot\rho} = \b{\ddot x} - \b{\ddot B}$.
The second derivative of the scalar $\rho(t)$ is then
\begin{equation}
  \ddot \rho = {d^2\over dt^2} (\b \rho \cdot \b \rho)^{1/2}
          = {\h \rho} \cdot \b{\ddot \rho} + \rho \, | \b{\dot{\hat\rho}} |^2,
\label{eq:ddotrho}
\end{equation}

i.e. the projection of the second derivative of ${\b \rho}$
onto the radial direction compensated by the centrifugal acceleration.
In order to determine $\rho(t)$ we need the physical
acceleration of observer and object as well as the angular velocity.

Although the orbit does not project to a perfect great circle as
viewed from the E-M barycenter, an approximate great circle fit
provides an adequate measure of the angular velocity required by
Eq~\ref{eq:ddotrho} because the progress along the great circle fit is
allowed to vary, anchored by the all the observations.
If the unit vectors from barycenter to
object $\h \rho_i$ have uncertainties $\delta\theta_i$ we can find a
great circle with pole $\h g$ that most closely passes through the
points.  Form a ``great circle merit function''
\begin{equation}
  \chi^2_g = \Sigma(\h g \cdot \h \rho_i)^2  (\delta \theta_i)^{-2} 
  \label{eq:chig}
\end{equation}
that sums the squared deviations of the points from the equator of this
great circle.  The least squares solution is
\begin{equation}
  0 = \begin{pmatrix}
       b_{xx} & b_{xy} & b_{xz} \\
       b_{yx} & b_{yy} & b_{yz} \\
       b_{zx} & b_{zy} & b_{zz} \\
  \end{pmatrix}
  \begin{pmatrix}
       g_x \\
       g_y \\
       g_z \\
  \end{pmatrix},
  \label{eq:gcpole}
\end{equation}
where $b_{jk} = \Sigma \hat \rho_j \hat \rho_k (\delta \theta_i)^{-2}$.
This is singular because it does not incorporate the constraint that
$|\h g|=1$.  Rather than imposing this via Lagrange multiplier or a
quadratic $(|\h g|^2-1)$ term
simply set $g_k=1$ for the $k$
with minimal $|b_{kk}|$, solve for the other two components, and
renormalize.  Because the $\delta\theta$ are so small this causes
negligible error.

Once the great circle pole is known, the azimuthal angle of each
observation around the great circle with respect to a reference
observation at $t_0$ can be estimated from the normals $\h n_i$ to the
planes defined by $\h g$ and $\h \rho_i$:
\begin{equation}
  \h n_i = {\h g \times \h \rho_i \over |\h g \times \h \rho_i|},
  \qquad
  \phi_i = \tan^{-1}(\h n_i \cdot \h g \times \h n_0, \;
                           \h n_i \cdot \h n_0).
\end{equation}
This set of $\phi_i$ is fitted with a polynomial in $t_i-t_0$,
and its derivative provides a measure of the instantaneous angular velocity.
The \PUMA\ library will use up to a cubic polynomial, given 4 or more
observations at diverse times, because this ``along track
acceleration'' can change rapidly when an asteroid ventures near the
Earth-Moon system.

If desired, the great circle fit to the unit vector $\h \rho$ at any
given time along the great circle may be calculated from the pole $\h g$,
the point on the great circle at the reference time
$\h h_0 = \h n_0 \times \h g$, and the angle $\phi(t_i-t_0)$,
\begin{equation}
  \h \rho = \cos\phi \; \h h_0 + \sin\phi \; (\h g \times \h h_0).
\end{equation}
and this can be compared with $\h \rho_i$.  The \PUMA\ library
makes this available to the user although it is not used otherwise.


This first iteration with a linear $\rho(t)$ now provides values for
$\ddot \rho_i$ and we iterate a second time with a quasi-cubic fit for
$\rho(t)$ that provides new predictions at the time of each
observation using a linear change in acceleration between the
reference time and each observation,
\begin{equation}
  \rho_i - \rho_0 = \dot \rho_0 (t_i-t_0) +
     {1\over3} (\ddot \rho_0 + {1\over2} \ddot \rho_i) (t_i-t_0)^2.
\label{eq:linrho}
\end{equation}
(For clarity we do not show the factor of $(1+\dot \rho_0 /c)$ that
needs to multiply $(\rho_i-\rho_0)$ to express the light travel time
$t_i^\prime-t_i$.)

For most purposes and for distances greater than 0.01~AU
Eq.~\ref{eq:linrho} suffices.  However, when an object comes very near
the Earth it is necessary to add a quadratic term to the distance
acceleration of Eq~\ref{eq:linrho}.  The \PUMA\ library implements
this (when there is sufficient time diversity) by fitting a quadratic
function to all the gravitational acceleration values, evaluating
$\ddot \rho$ as a quadratic function of time, and integrating it twice
as in Eq.~\ref{eq:linrho} to obtain each $\rho_i$ from a quartic
polynomial for $\rho(t)$.  Such a quadratic acceleration correction is
iterated twice.

Given this assumption for a distance and radial velocity $\rho_0,\dot\rho_0$,
the function $\rho(t)$ allows us to do a least-squares solution for the
state vector $(\b x_0, \b v_0)$ that describes the trajectory
that is most consistent with the observations.  We expect that
the object trajectory in the solar system is
\begin{equation}
  \b x(t) = \b x_0 + \b v_0 (t-t_0) + {1\over2} \b a_0 (t-t_0)^2 +
     {1\over6} \b a_1 (t-t_0)^3 + {1\over12} \b a_2 (t-t_0)^4,
\end{equation}
where the acceleration encountered along the trajectory is
\begin{equation*}
  \b a = \b a_0 + \b a_1 (t-t_0) + \b a_2 (t-t_0)^2.
\end{equation*}
($\b a_2=0$ if there are fewer than 3 time diverse observations.)

Note that the distances $\rho_i$ and ranges $r_i$ have already been
evaluated for this assumed $\rho_0,\dot\rho_0$, and the accurately
known $\h \rho_i$ implies that the acceleration coefficients, $\b a_0$,
$\b a_1$, and $\b a_2$, are already
accurately known for whatever best fit $\b x_0,\b v_0$ will emerge.
Since we can treat $\b a_0$, $\b a_1$, and $\b a_2$ as constant, the
predicted $\b x(t_i)$ are then linear functions of the state vector as
are the predicted observed unit vectors
$(\b x(t_i) - \b O^\prime_i)/r_i$
so these unit vectors can be compared with the observed unit vectors
through an appropriate inverse covariance matrix to form a $\chi^2$
merit function similar to Eq~\ref{eq:chig}.  
In the \PUMA\ library it is
advantageous to distinguish along-track versus cross-track errors
because these differ for streaked detections.  Setting the derivatives
of $\chi^2$ with respect to $(\b x_0, \b v_0)$ equal to zero provides
equations for the best fit $(\b x_0, \b v_0)$, and the minimum value
of $\chi^2$ provides an assessment of whether the assumed
$(\rho_0,\dot\rho_0)$ are a reasonable match to the data.

Of course finding the $(\b x_0, \b v_0)$ that best threads through the
observed angular positions, given that the radial distance and
velocity are essentially fixed, is really a 4D angular problem, but it
is computationally more efficient to solve the three, linear,
Cartesian position--velocity equations, not to mention avoiding
the singularity at the poles.

This solution at an assumed $(\rho_0,\dot\rho_0)$ differs slightly
from an exact orbital trajectory that could be obtained by classic
differential equation integration in two ways: the acceleration along
the trajectory is fitted as a first or second order polynomial, and
the acceleration is derived using a distance function that is also a
polynomial.  Even if the iteration to this solution converges, the
orbit and object locations at times $t_i$ are slightly different from
actuality.  Within the domain of validity for \PUMA, the difference is
much less than ground-based observational error, however.

Given $\chi^2(\rho_0,\dot\rho_0)$, a non-linear
least squares fit for a best fit $(\rho_0,\dot\rho_0)$ is
straightforward, although a few technical details are important.  The
fitted variables are actually $\ln(\rho_0)$ and $(\dot\rho_0/\rho_0)$ in
order to keep $\rho_0$ positive and avoid the large covariance between
$\rho_0$ and $\dot\rho_0$.  A grid search for initial conditions and
priors on $\rho_0$ and $\dot\rho_0$ can help prevent the best fit
solution from running away to unreasonable values.

The \PUMA\ library spends about 1 microsecond per observation
(involving computing observatory, Earth, Moon locations in the J2000
ecliptic coordinate system, etc), and $(\rho_0,\dot\rho_0)$
computation (involving the orbit fit iteration described above), so a
10$\times$10 grid initialization on 4 observations followed by a
Levenberg-Marquard least squares fit completes in less than a
millisecond.

The accuracy of the \PUMA\ extrapolation can be demonstrated using a
near Earth asteroid 2012~KF47 which passed by the Earth near the end
of Feb 2022.  ATLAS observed this NEO from all four observatories for
two weeks during which its distance $\rho$ changed from 0.24~AU to
0.29~AU, its radial velocity component $\dot\rho$ changed from
+1.8~km/s to +6.9~km/s, and its tangential velocity changed from
1.30~deg/day to 1.33~deg/day to 1.11~deg/day.

Downloading the exact ephemerides for 2012~KF47 from JPL
Horizons\footnote{\tt https://ssd.jpl.nasa.gov/horizons}, we
do a \PUMA\ fit to a short arc of 4 RA and Dec points spanning 0.9
hour on MJD~59609 using the precise JPL Horizons RA and Dec
coordinates.  (We chose this particular MJD because the angular
velocity has a substantial acceleration at that moment of the
passage.)  The result is that the error in the \PUMA\ calculated RA
grows quadratically with time at a rate of $-9$~arcsec/day$^2$ and the
Dec error is $+12$~arcsec/day$^2$.  The \PUMA\ derived distance is
smaller than true distance by 2.6\%, and the fitted radial velocity is
32\% smaller than the true velocity.

If we instead present \PUMA\ with 4 points that span the same 0.9
hour on MJD~59609 and an additional 4 points spanning 0.9 hour one
day later, the \PUMA\ fit differs from JPL Horizons by only
0.03~arcsec in RA and 0.01~arcsec in Dec after a span of 20 days.  The
\PUMA\ fit distance is 0.5\% smaller than the true distance and the radial
velocity is 16\% larger than the true velocity.

The reason for the improvement with 8 points over a longer time span
is that \PUMA's polynomial approximations to the gravitational
acceleration and angular motion can be misled when \PUMA\ only sees a
very short arc for an object such as 2012~KF47 which in fact has
substantial angular acceleration.  Based on a time span of only
0.9~hour, the angular velocity that goes into Eq~\ref{eq:ddotrho} is
unrepresentative enough that the centrifugal contribution to
$\ddot\rho$ creates this quadratically growing error in the
along-track extrapolation.  When \PUMA\ is given 8 points that span a
$\delta t$ that is about 50 times longer the error from the
centrifugal contribution becomes negligible.  For any asteroid whose
angular velocity is changing linearly with time the 0.9~hour arc would
have been sufficient to achieve this $\sim$30~milliarcsec accuracy.

This particular \PUMA\ calculation took 0.4~millisec for 8 points, if
we dispense with an initial grid search, including the non-linear
least squares fit for the distance and radial velocity, or 3~millisec
with 8 points if we ask for a full grid search to be sure of good
initial conditions (the result is the same).

During the presentation below of \PUMALINK\ we describe more
examples of \PUMA\ performance on very nearby asteroids, and we
distribute other examples with our source code, including an impactor
and the Tesla roadster.  The \PUMALINK\ tests run \PUMA\ on
billions of cases, and the resulting $\chi_\nu^2$ values for real
objects demonstrate that \PUMA\ robustly produces orbits consistent
with observations for a very wide variety of input data.

Two other orbit fitters in common use are ``{\tt openorb}''
\citep{2009M&PS...44.1853G} and Bill Gray's ``{\tt
  Find\_Orb}''\footnote{\tt
  https://www.projectpluto.com/find\_orb.htm}.  Using each of these to
fit an orbit to these same points took about 10~sec for {\tt openorb}
and 0.1~sec to 0.3~sec for {\tt Find\_Orb}.  These two packages are
more sophisticated than the \PUMA\ library in terms of number of
perturbers in the Solar System, integration accuracy, allowable span
between data points, etc, but the \PUMA\ library excels at raw speed
with enough accuracy for any linking problem.

It is important to keep in mind that the \PUMA\ library is fitting the
points it is given, and its interpolation among these points is
accurate at the 10's of milliarcsec level.  Although \PUMA\ is
remarkably accurate when asked to extrapolate beyond its data points,
it is evaluating a shrewdly chosen set of polynomials and not actually
doing an exact differential equation integration with the usual
acceleration evaluations and predictor-corrector iterations.
Therefore \PUMA's accuracy will degrade with extrapolation times
$\Delta t$ that are more than $\sim$1000 the inter-observation time
$\delta t$ or extrapolations over a time span during which the Earth's
orbit deviates significantly from a polynomial.  On the other hand,
\PUMA\ is explicitly using the gravity of the Sun, Earth, and Moon, so
its accuracy does not erode because an object is near the Earth.

\section{Linking}

The previous section illustrates how to determine a distance and
radial velocity from three or more observations, or, given a distance
and radial velocity, how to project a pair of observations to a
different ephemeris time.  The object's motion relative to the
Earth-Moon barycenter is smooth and well
described by a polynomial approximation for the acceleration it feels,
so the results returned by \PUMA\ are accurate at the observational
uncertainty level for at least a month.

In this section we will examine how to decide whether linked pairs or
sets of observations (``tracklets'') taken at very different times
should be tested against one another with \PUMA.  Since it takes of
order 1 millisecond to do an accurate \PUMA\ test we easily can afford to
perform $10^5$ \PUMA\ tests per observation, but $10^7$ or $10^8$
tests becomes excessively expensive so we cannot blindly test all
pairs of tracklets.  Our goal is to find a way to determine whether
two tracklets {\it cannot} link, with a compute time less than 1
microsecond, so the remaining possibilities can be passed on to
\PUMA\ for a more rigorous test.

There are two sources of uncertainty when comparing two tracklets at
very different times.  The first is simply the statistical uncertainty
that arises from the astrometric uncertainty of each detection.  This
is typically a fraction of an arcsecond at the moment of detection,
depending on SNR, but of course it grows with extrapolation in time.
The second uncertainty arises because the detections that comprise a
tracklet could be at any distance and radial velocity.


In this section we will examine how the statistical uncertainty
propagates in time, then the effects of distance and radial velocity
uncertainty, and finally a fast method to compare pairs of tracklets.

\subsection{Propagation of measurement uncertainty}

Given a linked pair of detections, and we want to know where an
object creating these detections will be found at a different time.
Given coordinates $\b p_1$ and $\b p_2$ with uncertainties
$\sigma_1$ and $\sigma_2$, separated by a time interval $\delta t$, we
assign an initial position and velocity
\begin{align*}
  \b p_0 &= [\b p_1/\sigma_1^2+\b p_2/\sigma_2^2] \;
            [1/\sigma_1^2+1/\sigma_2^2]^{-1}\\
  \b v_0 &= (\b p_2-\b p_1)/\delta t.
\end{align*}

The uncertainty in the tracklet components is
\begin{align*}
  \langle p_0^2\rangle &= \left[1/\sigma_1^2+1/\sigma_2^2\right]^{-1} \\
  \langle p_0 v_0\rangle &= 0 \\
  \langle v_0^2\rangle &= (\sigma_1^2+\sigma_2^2)/\delta t^2.
\end{align*}
assuming $\b p_1$ and $\b p_2$ are uncorrelated.
(The averaging indicated by $\langle\rangle$ is understood to be
relative to the mean values.)

Assuming unaccelerated evolution, propagating this tracklet by time
$\Delta t$ gives
\begin{align*}
  \b p(t) &= \b p_0 + \b v_0 \; \Delta t \\
  \b v(t) &= \b v_0.
\end{align*}

The uncertainty in the tracklet after an
interval $\Delta t$ is
\begin{align*}
  \langle p^2\rangle &= \langle p_0^2\rangle + \langle v_0^2\rangle \Delta t^2
       = \left[{1/\sigma_1^2}+{1/\sigma_2^2}\right]^{-1}
       + (\sigma_1^2+\sigma_2^2)\Delta t^2/\delta t^2 \\
  \langle p v\rangle &= \langle v_0^2\rangle \Delta t = 
      (\sigma_1^2+\sigma_2^2)\Delta t/\delta t^2 \\
  \langle v^2\rangle &= \langle v_0^2\rangle =
      (\sigma_1^2+\sigma_2^2)/\delta t^2.
\end{align*}

As an illustration, if $\sigma_1=\sigma_2=\sigma$ we get for a covariance
matrix for $p,v$ after interval $\Delta t \gg \delta t$
\begin{equation}
  \b C = {2 \sigma^2\over \delta t^2} \begin{pmatrix*}
    \Delta t^2 + \delta t^2/4& \quad \Delta t \\
    \Delta t     & \quad 1 \\
  \end{pmatrix*},
\label{eq:cov}
\end{equation}
and inverse covariance matrix
\begin{equation}
  \b C^{-1} = {2 \over \sigma^2} \begin{pmatrix*}
    1 & \quad -\Delta t \\
    -\Delta t  &  \quad \Delta t^2+\delta t^2/4 \\
  \end{pmatrix*}.
\label{eq:invcov}
\end{equation}
Note that the determinant of $\b C$ is completely independent of time
$\Delta t$: the $p,v$ astrometric uncertainty volume just squeezes out
into an ever thinner line even though the uncertainty
on $p$ individually grows linearly with $\Delta t$.

If $\b p$ comprises two uncorrelated coordinates such as unit tracklet
components or angle on the sky, the 6D or 4D covariance matrix
consists of $2\times2$ blocks for the displacement and velocity in
each coordinate.  If the uncertainties in these coordinates are
correlated, for example because the along-track error is different
from the cross-track error, the covariance matrix is more complicated.

\subsection{Influence of $\rho,\dot\rho$ on tracklet propagation}

Suppose for the moment that an object has no acceleration relative to
the usual Earth-Moon barycenter origin and moves at constant velocity.
Two angular observations on the sky, augmented by an assumption about
distance and radial velocity, allow us to calculate the trajectory of
this object.

Referring again to Fig~\ref{fig:pumavar}, suppose we have two
observations at $t_0$ and $t_1$ (dropping the primes for clarity) with
time interval $\delta t = t_1-t_0$, of an object that is moving at
constant velocity relative to the Earth-Moon barycenter.  We want to
know where it is after at a reference time $\Delta t$ after $t_0$.  If
we assume that the object has a distance and radial velocity of
$\rho,\dot\rho$ at this reference time, we can linearly extrapolate
the two earlier observations to get the 3D object position $\b x$
relative to the barycenter origin.

We divide by this result by $\rho$ to get the extrapolated unit vector
at that time.  To second order in $1/\rho$, and taking $\delta
t\ll\Delta t$,
\begin{align}
  {1\over\rho}\;\b x(\Delta t) &\approx \b{\hat u}_0 + \Delta t\,\b{\dot\varphi}
  \nonumber \\
  &+ {1\over\rho}\left[
    -\dot\rho\,\Delta t^2\,\b{\dot\varphi} + \b{\dot D}\,\Delta t +\b D_0^\prime
    - D_\parallel\,(\b{\hat u}_0 + \Delta t\,\b{\dot\varphi})
    \right]
  \nonumber \\
   &- {1\over2\rho^2}\left[
      (D_\perp^2+\dot\rho^2\Delta t^2)\,\Delta t\,\b{\dot\varphi}
    + (D_\perp^2-\dot\rho^2\Delta t^2)\,\b{\hat u}_0)
    \right],
\label{eq:linextrap}
\end{align}
where the observed tangential motion of the object and observatory
motion are defined as
\begin{equation*}
  \b{\dot\varphi} \equiv {1\over\delta t}(\b{\hat u}_1-\b{\hat u}_0), \quad
  \b{\dot D} \equiv {1\over\delta t}(\b D_1^\prime-\b D_0^\prime),
\end{equation*}
and  $D_\perp$ and $D_\parallel$ are
the perpendicular and parallel components of $\b D_0^\prime$ with
respect to $\b{\hat u}_0$ as above.
Equation~\ref{eq:linextrap} may be differentiated with respect to
$\Delta t$ to get the extrapolated angular velocity.

Note that if $D\sim R\earth$ and $\rho\sim1$~AU, $D/\rho$ is
${\sim}4\times10^{-5}$, $\dot D\Delta t/\rho$ is
${\sim}2\times10^{-3}$ over $\Delta t\sim$~10 days, and $\dot \rho$ is
an order of magnitude bigger than $\dot D$ for typical orbital
velocities.  Therefore the terms involving $D$ make a small
contribution in Eq~\ref{eq:linextrap}.

This is the motivation for computing the motion relative to the
Earth-Moon barycenter, as close to the observations as possible
without incurring rapid wiggles in inertial space.  HelioLinC, using
a the Sun as the origin with $D\sim1$~AU, computes a motion that is quite
sensitive to the choice of the (unknown) $\rho,\dot\rho$, and
the following analytic progress is impossible unless $\rho\gtrsim1$~AU.

Define $s$ as the inverse distance,
and $w$ as the inverse of the ``collision time'', before a radially
moving object will arrive at the Earth:
\begin{equation*}
  s \equiv {1\over\rho} \quad
  \hbox{and} \quad
  w \equiv {\dot\rho\over\rho}\;.
\end{equation*}
(These variables $s$ and $w$ are almost exactly the same as $\gamma$
and $\dot\gamma$ used by \cite{2000AJ....120.3323B} and
\cite{2018AJ....156..135H}, but the different origins make the
variables slightly different.  These origin differences are in fact
critical, the use of dimensional $\gamma$ and $\dot\gamma$ along with
their dimensionless $\alpha$ and $\beta$ can be confusing, and
\PUMA\ explicitly avoids spherical coordinates.  We deem it advisable
therefore not to reuse these variable names from the literature.)

Neglecting terms proportional to $D$, we can rewrite
Equation~\ref{eq:linextrap} in terms of the variables $s$ and $w$,
\begin{equation}
  {1\over\rho}\;\b x(\Delta t) \approx
     \left[ {3\over2} - {(1+w\Delta t)^2\over 2}\right]
            \Delta t\,\b{\dot\varphi} + 
     \left[1+{(w\Delta t)^2\over 2}\right] \b{\hat u}_0
     + \b{\dot D}\,\Delta t\,s.
\end{equation}
The term $w\Delta t$, the ratio between extrapolation time and
collision time, is typically small unless impact is imminent, so its
square is smaller still:
\begin{equation}
  w\Delta t  = {\dot\rho \over \rho}\,\Delta t = 0.06
       \left( {\dot\rho\over 10\,\hbox{km/s}} \right)
       \left( {\rho\over 0.1 \hbox{AU}} \right)^{-1}
       \left( {\Delta t\over 1\hbox{day}} \right).
\end{equation}

Of course a proper \PUMA\ extrapolation neglects neither $D$ nor the
accelerated motion of the object with respect to the barycenter, but
these only modify the exact polynomial coefficients and do not
fundamentally change the linear behavior for $|w\Delta t|\ll1$.

This implies that each component of the extrapolated unit vector and
extrapolated velocity divided by distance from a pair of observations
spans a triangle in 6D space, with vertices at $\rho=\infty$ and the
minimum and maximum plausible $\dot\rho$ at the minimum $\rho$
considered.  Using \PUMA\ to accurately extrapolate the observations
with a test set of $\rho_0,\dot\rho_0$ to the reference time, a planar
fit to each of these coordinates as a function of the resulting $s,w$
is very accurate when $|w\Delta t|\ll1$.

That the manifold of possible extrapolations in 6D phase space for a
pair of observations is 2 dimensional is not surprising, given that it
is determined by the unknown $\rho,\dot\rho$, but it is a nice
surprise that it is a nearly planar triangle and nearly linear in
$s,w$.  The size of the triangle and extent of uncertainty depends on
the minimum value of $\rho$ and maximum absolute $\dot\rho$ according
to $|w\Delta t|$, so the computer resources can be tailored according
to how close the user wants to approach impact time.
Section~\ref{sec:neoex} provides some illustration of these triangles

Venturing to very small $\rho$ will increase $w$, make $|w\Delta t|\sim 1$,
and cause non-linear terms such as $w^2$, $sw$, and $s^2$
to become significant.  (Although the position does not have $sw$ and
$s^2$ terms, the velocity does.)  By restricting $w\Delta t$ to the
linear regime, we can exploit this approximation to test the linkage
between observations very quickly over a wide range of
$\rho,\dot\rho$.

\subsection{Linking two pairs of observation}

We now have the ingredients to test whether two pairs of observations
taken at time $t_1$ and $t_2$ might link with one another.  We start
by picking a common reference time when they can be compared.  Because
there are terms that depend on $\Delta t^2$ and because the validity
of the linear approximation of the previous section depends on
$|w\Delta t|\ll1$ it is advantageous to choose an comparison time that
makes $|\Delta t_1|$ and $|\Delta t_2|$ as small as possible, for
example half way between $t_1$ and $t_2$.

The results of the previous section allow us to calculate the unit
vector and distance divided velocity of each of these tracklets at this
reference time, expressing the result as a polynomial in the unknown
$s,w$ distance and radial velocity at the reference time.

For a choice of $\rho,\dot\rho$ (i.e. $s,w$) designate $\b S_1(s,w)$ and $\b
S_2(s,w)$ as the distance-divided 6D state vectors at the reference
time for the two tracklets: these are each polynomials in $s,w$ where
the fit to the trial values propagated using \PUMA\ gives us all the
coefficients.  We can also assemble the 6D covariance matrices $\b C_1$
and $\b C_2$, which we take to comprise three 2$\times$2 blocks for
each coordinate as described in Eq~\ref{eq:cov} above.  We assume that
$\b C_1$ and $\b C_2$ do not depend on $s,w$, but we do augment them
by the covariance matrix from the errors of the linear fit with
respect to the grid of points that produce the coefficients for $\b
S_1$ and $\b S_2$.

Then the simultaneous statistical compatibility of some arbitrary state
vector $\b S$ with $\b S_1(s,w)$ and $\b S_2(s,w)$ is governed by
\begin{equation}
  \chi^2(s,w) = (\b S - \b S_1)^T\,\b C_1^{-1} (\b S - \b S_1) +
   (\b S - \b S_2)^T\,\b C_2^{-1} (\b S - \b S_2).
  \label{eq:chi}
\end{equation}
We are seeking a solution such that $\b S_1$ and $\b S_2$ are
evaluated at the same $s,w$; therefore since $\b S_1$ and $\b S_2$ are
polynomials of $s,w$, $\chi^2(s,w)$ is also a polynomial of $s,w$.

Taking the derivative with respect to $\b S$, the minimum of $\chi^2$
occurs at the inverse variance weighted average of the two locations
\begin{equation}
  \b S_{min}(s,w) = \b C \left( \b C_1^{-1} \b S_1 +  \b C_2^{-1} \b S_2 \right),
  \label{eq:xmin}
\end{equation}
where the combined inverse covariance matrix is the sum of the two
matrices from the two tracklets
\begin{equation}
  \b C^{-1} = \b C_1^{-1} + \b C_2^{-1}.
\end{equation}

Eq~\ref{eq:xmin} is linear in $\b S_1(s,w)$ and $\b S_2(s,w)$, so the
most statistically likely $\b S_{min}(s,w)$ is also a polynomial in
$s,w$ of the same order.  Plugging this result back into
Eq~\ref{eq:chi} yields an expression for $\chi^2$, which has
polynomial terms from the matrix squares of $(\b S_{min}-\b S_1)$ and
$(\b S_{min}-\b S_2)$.  This expression for $\chi^2$ is therefore now
quadratic in $s,w$, with coefficients multiplying 1,  $s$ , $w$, $w^2$, $sw$, and $s^2$.  (A quadratic fit to $\b S_1,\b S_2$
leads to a quartic $\chi^2$ and 15 coefficients.)

An expression for $\chi^2(s,w)$ that is quadratic in $s,w$ can be cast
in the form:
\begin{equation}
  \chi^2(s,w) = \chi^2_0 + e_1(s-s_0)^2 + e_2(s-s_0)(w-w_0) + e_3(w-w_0)^2.
  \label{eq:chiellipse}
\end{equation}
The statistically most compatible distance and radial velocity are
evident as $s_0,w_0$, and the minimum value of $\chi^2(s,w)$ emerges
as $\chi^2_0$.  The contours in $\chi^2$ are nested ellipses whose
size and orientation is given by the coefficients $e_1$, $e_2$, and
$e_3$, and these can be converted to uncertainties and covariance in
$s_0$ and $w_0$.


Note that if we kept the quadratic terms for the fits to $\b
S_1$ and $\b S_2$ it would not fundamentally change this approach.
Finding the best fit $\b S_{min}$ and evaluating the $s,w$ that
minimize $\chi^2$ simply involves finding roots of higher order
polynomials, or more realistically, treating the quadratic terms as
perturbations on the linear function and adjusting the result
accordingly.  The current implementation is sufficiently accurate
without the quadratic terms.

\subsection{A NEO example}
\label{sec:neoex}

Returning to the NEO 2012~KF47 that ATLAS observed in Feb 2022 at
$\rho$ of 0.24~AU to 0.29~AU and $\dot\rho$ of +1.8~km/s to +6.9~km/s,
we can examine how \PUMALINK\ uses a pair of detections from two
different nights to determine whether the two tracklets link with one
another.

\begin{figure}[h]
\begin{center}$
\begin{array}{ccc}
\includegraphics[width=2.5in]{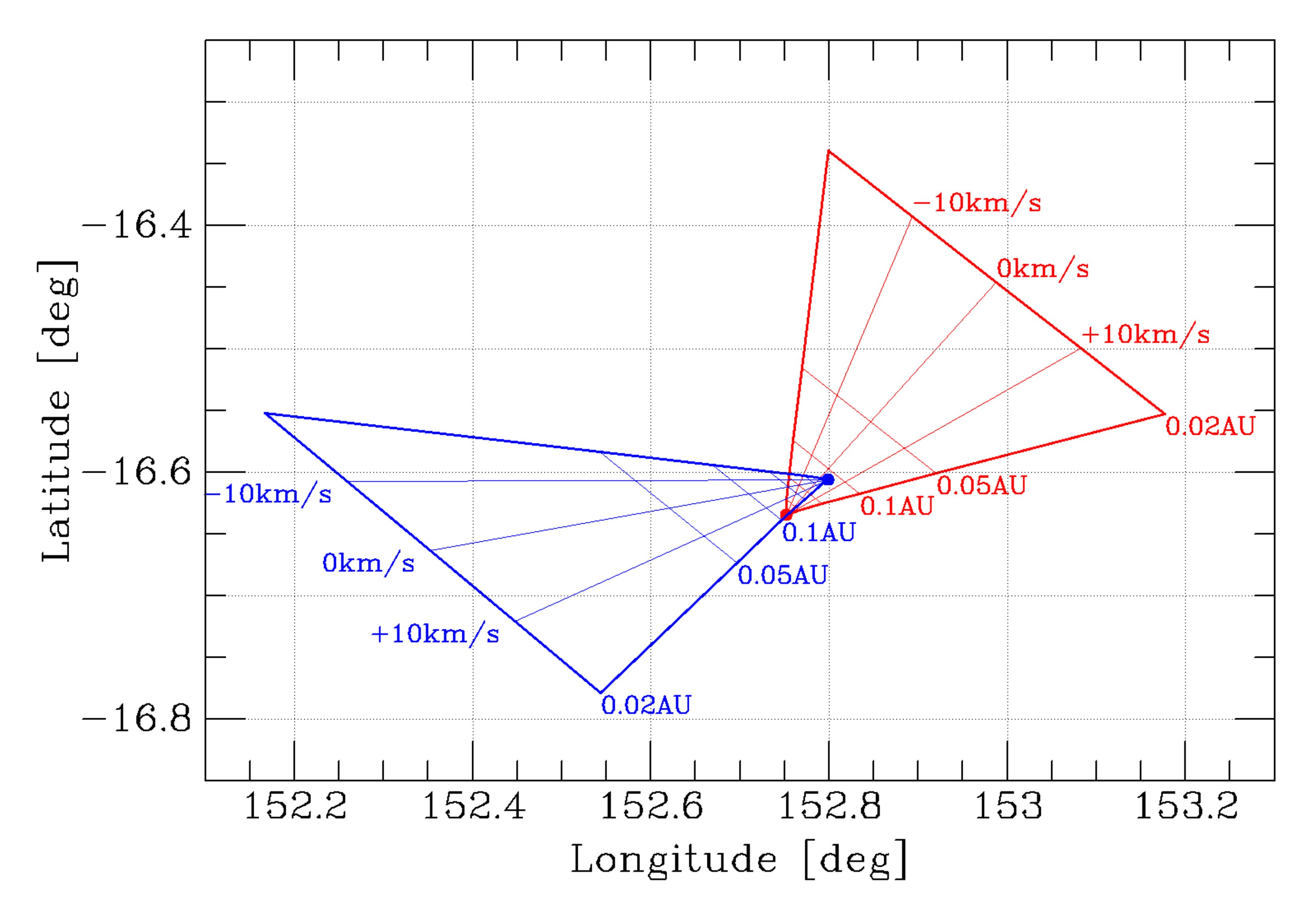}
\qquad
\includegraphics[width=2.5in]{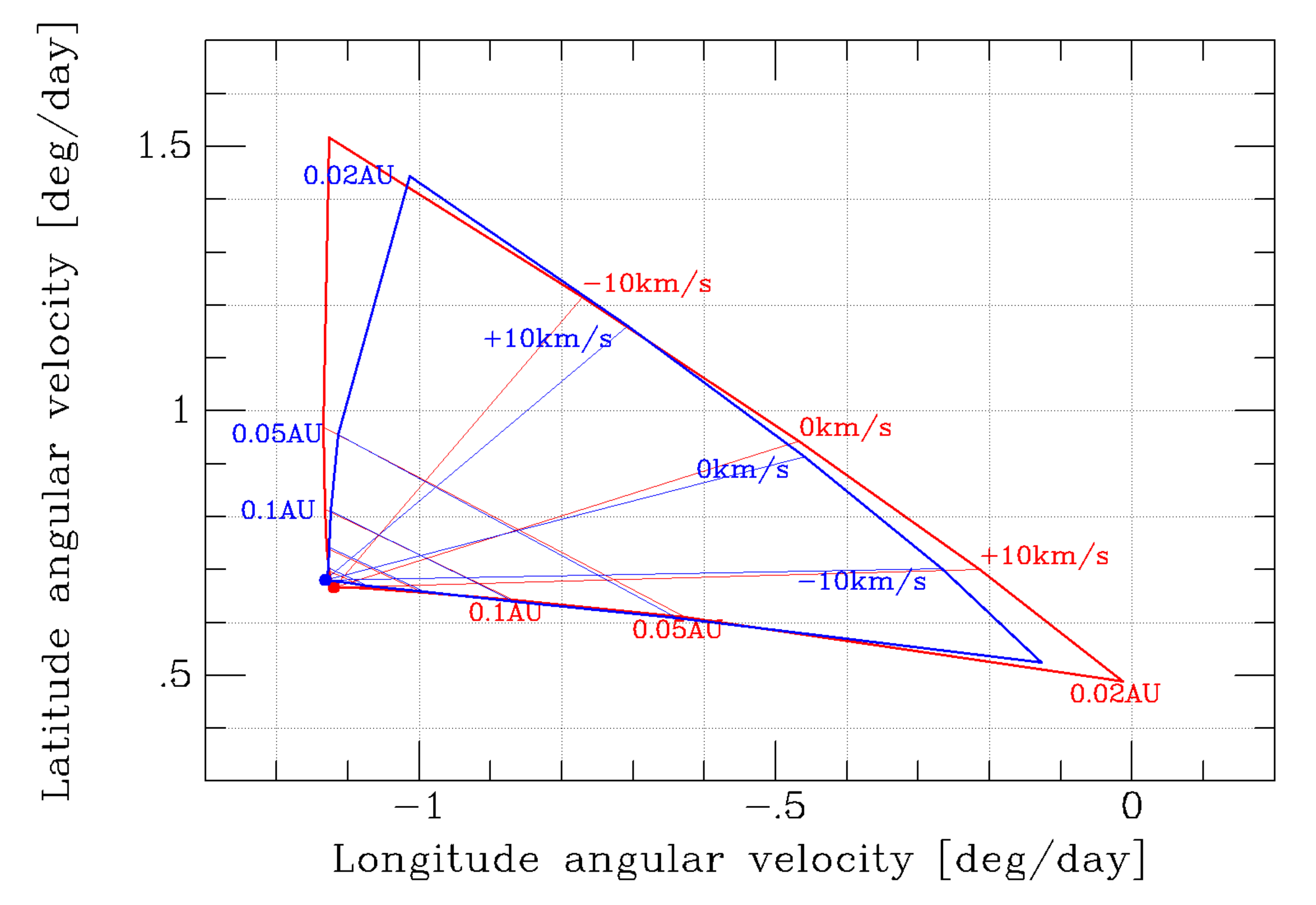}
\end{array}$
\end{center}
\caption{The NEO 2012~KF47 was observed by ATLAS on MJD 59606.47 and
  59607.53.  The left panel shows the extrapolated barycenter position
  for pairs of detections from each night at the intermediate MJD
  59607.0 for a variety of assumptions for $\rho,\dot\rho$ at that
  reference time, and the right panel shows the extrapolated angular
  velocity.  The extrapolation of the pair from the first epoch
  forward is shown in red, the extrapolation of the pair from the
  second epoch backward is shown in blue, and labels show selected
  values of $\rho,\dot\rho$.  A viable solution can only exist for an
  overlap in these four coordinates at a common value of
  $\rho,\dot\rho$.  The dots on one vertex of the triangles show the
  great circle extrapolation for $\rho$ infinitely far away.
}
\label{fig:trilap}
\end{figure}

Figure~\ref{fig:trilap} shows the \PUMA\ prediction for where this
NEO would be, as seen from the Earth-Moon barycenter, at MJD 59607.0,
given tracklets taken from 59606.47 and 59607.53.
This calculation
shows only the effect of $\rho,\dot\rho$ without taking observational
errors into account.  The triangle vertices near ecliptic longitude
152.8 and latitude $-16.6$ that correspond to great circle
extrapolation are offset by about 0.05 degree because the
object is in fact not infinitely far away; it is 0.242~AU distant at that time.
These two tracklets are consistent with one another if they match in
all four dimensions at the same $s,w$.

\begin{figure}[h]
\begin{center}$
\begin{array}{ccc}
\includegraphics[width=2.5in]{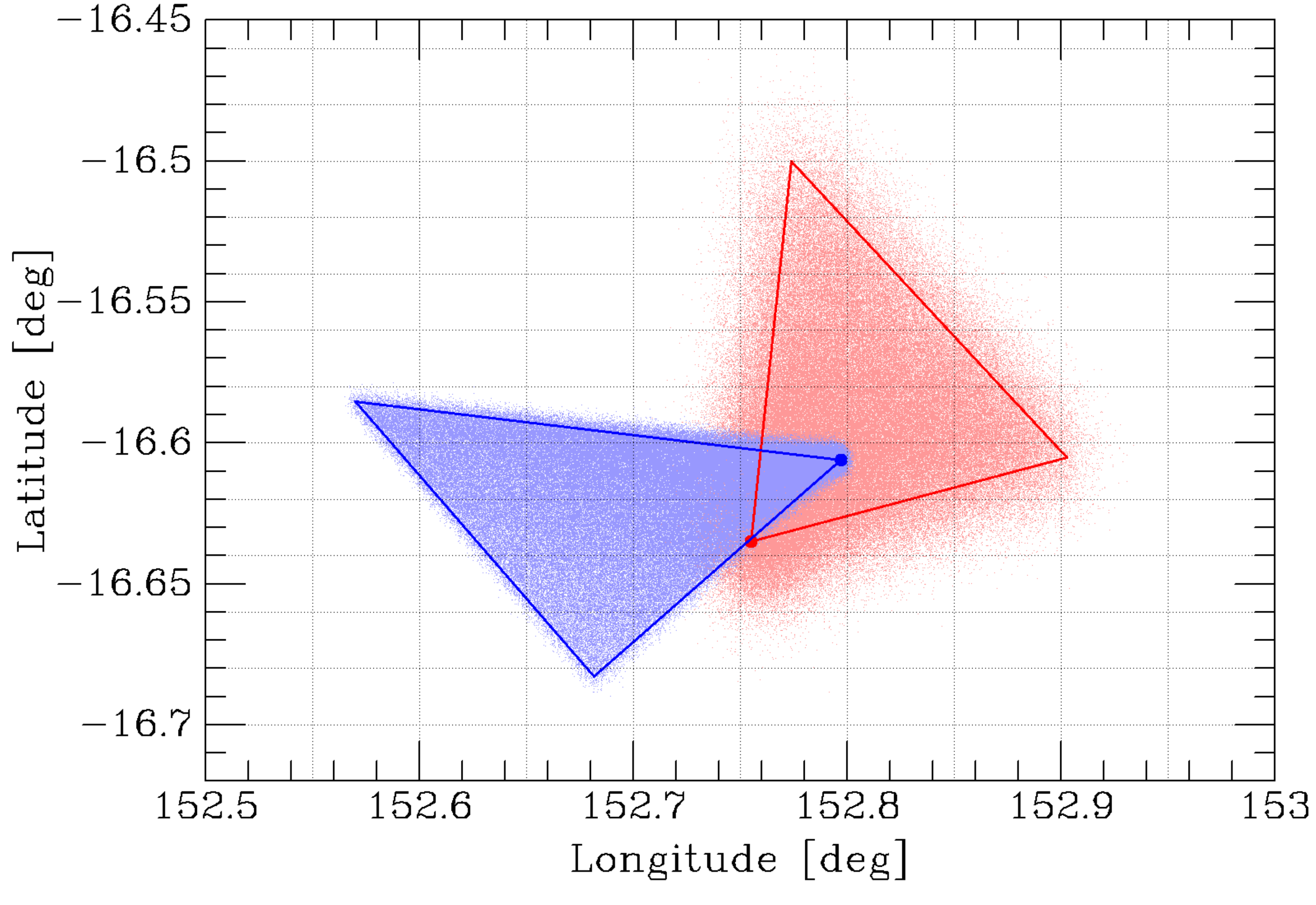}
\qquad
\includegraphics[width=2.5in]{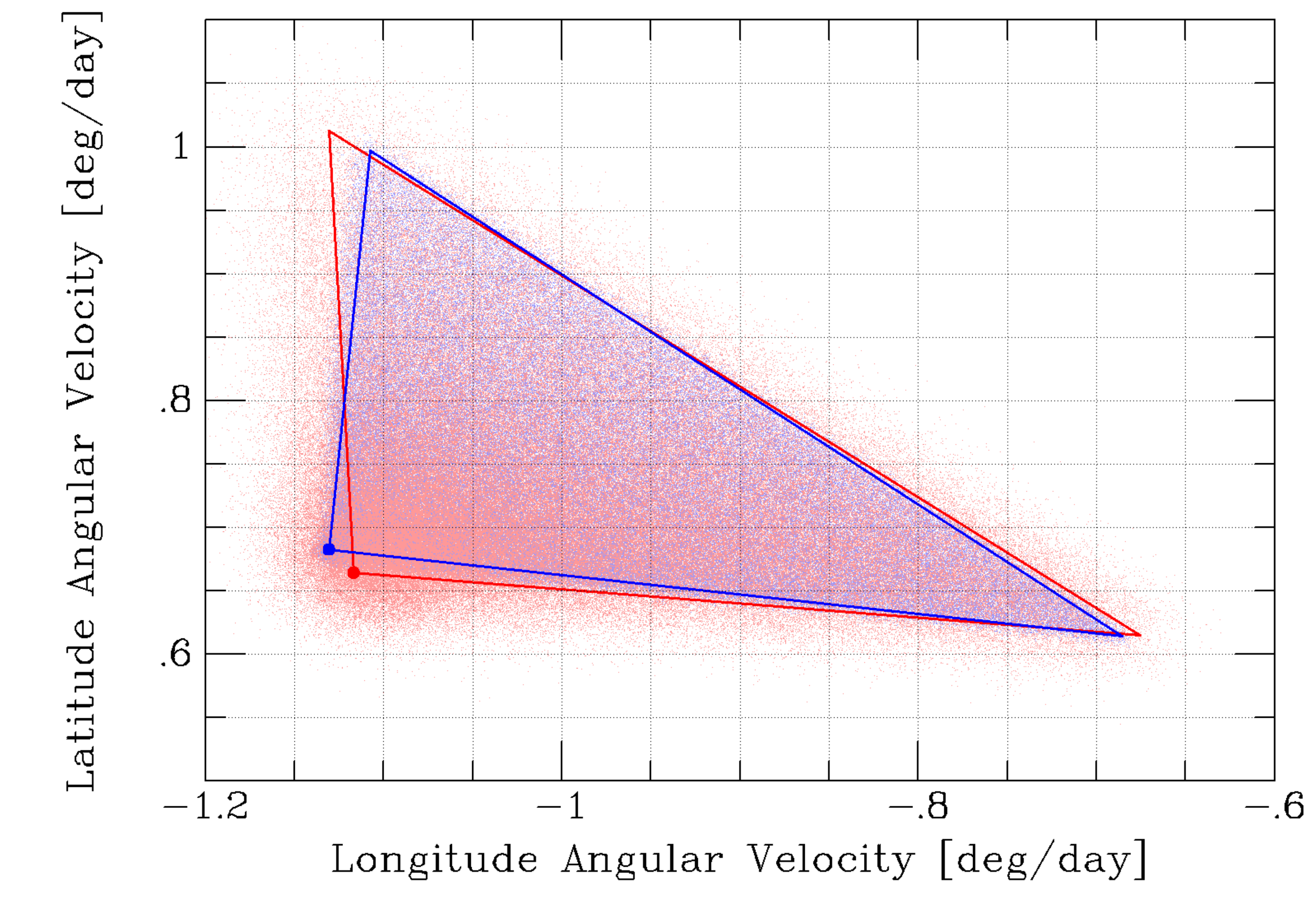}
\end{array}$
\end{center}
\caption{This figure shows similar calculations for NEO 2012~KF47 as
  Fig~\ref{fig:trilap} but expanded to show only a minimum distance of 0.05~AU
  and $\pm20$~km/s at the observational epochs, this time
  including random variations in all four observations.  The left
  panel shows the effect in position--position space, the right is
  velocity--velocity space.  The point at infinity is marked with a dot
  on the bounding triangles.  The scatter for the first (red) epoch is
  greater because $\delta t$ is 834 sec, compared with 3978 sec for
  the second (blue) epoch.}
\label{fig:randpp}
\end{figure}

Figure~\ref{fig:randpp} shows a subregion from Fig~\ref{fig:trilap}
%
with the scatter possible from observational uncertainties applied as
well.  This fuzzes out the triangles, and it implies that the two
tracklets do not need to match exactly at the same $s,w$ in order to be
statistically consistent.  Note that the uncertainty in where the
asteroid lies caused by uncertainty in $\rho,\dot\rho$ is considerably
bigger than the astrometric uncertainty at distances smaller than 0.2~AU.

\begin{figure}[ht]
\begin{center}
\includegraphics[width=3in]{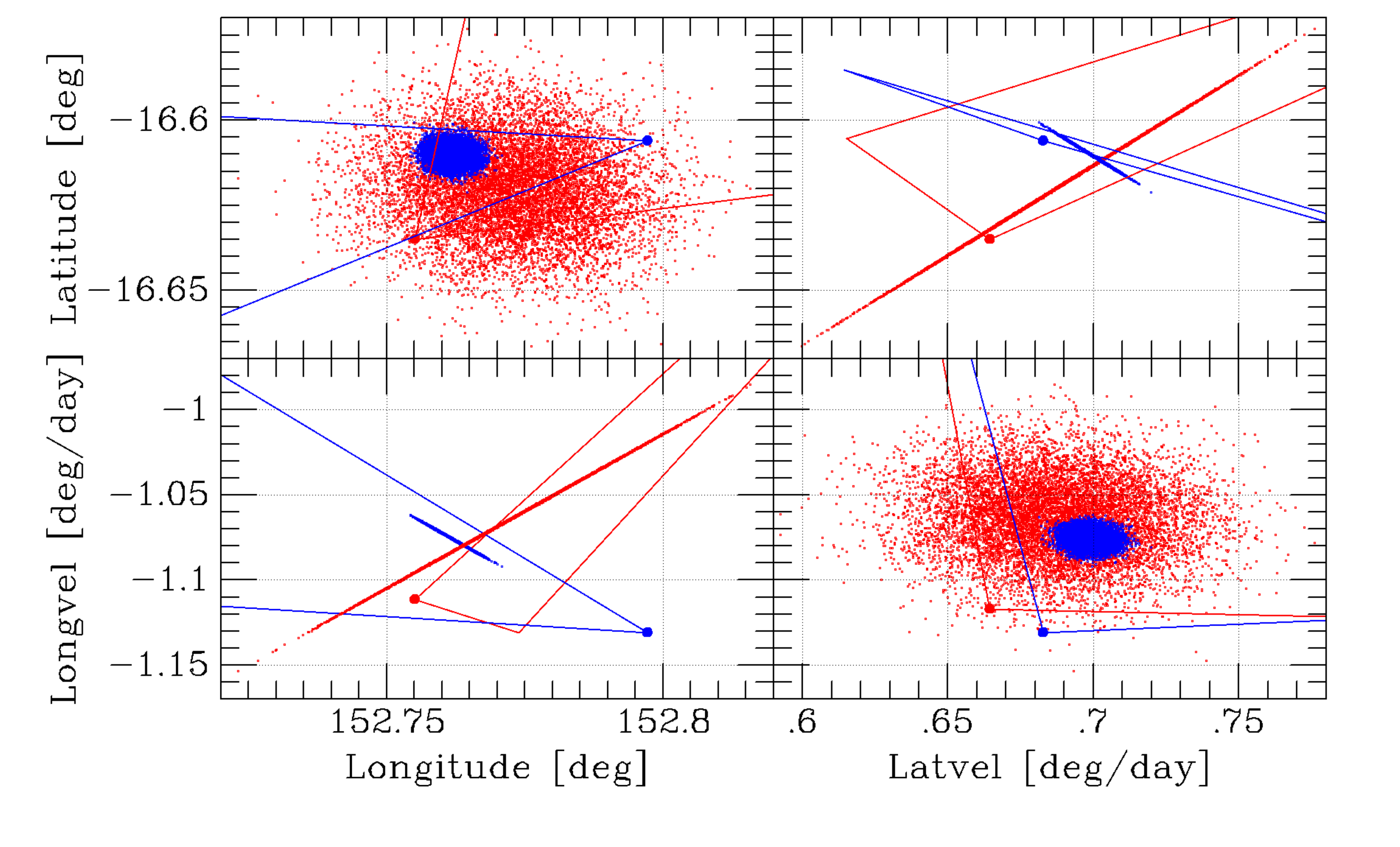}
\end{center}
\caption{The triangles that show the range of possibility for the
  location of the NEO 2012~KF47 caused by variation in $\rho,\dot\rho$ are
  expanded again from Fig~\ref{fig:randpp}.  In this figure the scatter caused
  by error in observational error is restricted to the correct $\rho,\dot\rho$ of
  0.242~AU, +2.04~km/s at 59607.0.  The upper left
  panel shows the scatter in the position--position plane, the lower left
  panel is longitude position--longitude velocity (latitude is in the
  upper right), and the lower right panel is the velocity--velocity plane
  (axes flipped with respect to Fig~\ref{fig:randpp}).
  As before the two epochs are indicated by red and blue.  At a fixed
  $\rho,\dot\rho$ the 4D observational scatter is a very thin disk
  that has a significant spread in position--position and
  velocity--velocity, but whose thickness in the position--velocity
  planes is comparable to the observational astrometric uncertainty.}
\label{fig:realpp}
\end{figure}

Figure~\ref{fig:realpp} repeats figure Fig~\ref{fig:randpp}, 
but statistically scattered points are only
shown for a particular $\rho,\dot\rho$, set to be the actual
values for 2012~KF47.

Recall that the 4D uncertainty volume is actually very small, however.
Figure~\ref{fig:realpp} demonstrates that although consideration of
only a positional match or velocity match bears a lot of uncertainty,
a proper consideration of the covariance between position and velocity
in each coordinate is very restrictive on whether two tracklets are
statistically consistent with one another.
The linear fit as a function of $s,w$ to the points seen in
Fig~\ref{fig:randpp} has an RMS error of 2.5~arcsec in angular
coordinate and 12~arcsec/day in angular velocity

\subsection{Software implementation}
\label{sec:sw}

\begin{figure}[ht]
\begin{center}
\includegraphics[height=3in]{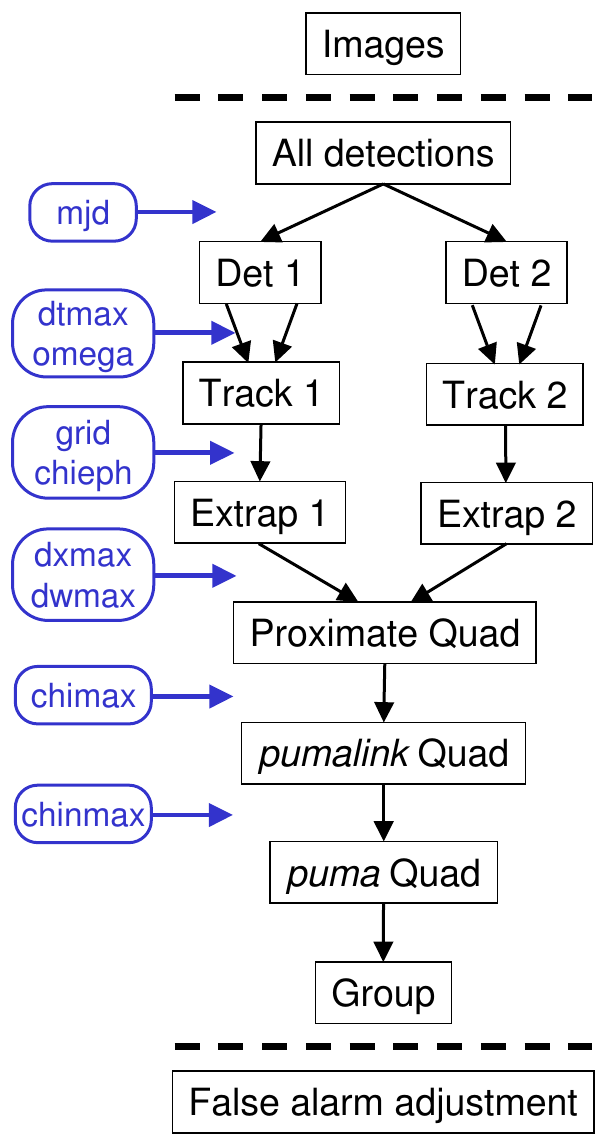}
\end{center}
\caption{The processing flow of \PUMALINK\ starts after a set of
  images have been processed to produce a set of detections, and after
  \PUMALINK\ completes the user must choose where to set the thresholds
  for detection probability versus false alarm rate.}
\label{fig:plflow}
\end{figure}

We have implemented these algorithms in a program called
``\PUMALINK''.  The processing follows a number of steps, illustrated
in Fig~\ref{fig:plflow}, and detailed below.
The only requirement for the user is to organize a set of detections
into ``{\tt TRD9}'' format, which is MJD, RA, Dec, cross-track
astrometric error, along-track astrometric error, observatory
longitude, latitude, and elevation, and an arbitrary, unique string
ID.  \PUMALINK\ works by organizing pairs of detections within an
epoch into tracklets and then linking pairs of tracklets from two
epochs into a quad.  The user should be mindful of this to ensure that
\PUMALINK\ will join detections into tracklets and identify the two
different epochs properly.

These are the \PUMALINK\ steps.
\begin{itemize}
  \item{} Read a set of detections in ``{\tt TRD9}'' format.
  \item{} Link these detections into tracklets, using different
    options: by default \PUMALINK\ internally links pairs of
    detections within a limiting time ({\tt dtmax}, default 0.1~day)
    and angular velocity ({\tt omega}, default 5~deg/day), but
    \PUMALINK\ can also use externally provided tracklets designated
    using the detection IDs.
  \item{} Use the \PUMA\ library to extrapolate each tracklet to the
    reference time ({\tt mjd}, normally taken to be the midpoint of the biggest
    time gap between detection times).  This is computed for a grid of
    possible distances and radial velocities, by default 5 steps in
    distance from 0.02~AU to 4~AU and 5 steps in velocity from
    $-20$~km/s to +20~km/s.  The results in each of the 6 unit vector
    and distance divided velocity components are fitted as a linear
    function of $s,w$, saving the coefficients and covariance matrix
    between each component's position and velocity with respect to the fit.
  \item{} Use a 6D kd-tree to find all pairs of tracklets that are
    sufficiently close at the reference time to bear further
    examination, nominally with search dimensions of {\tt dxmax} 0.2~deg and
    {\tt dwmax} 0.5~deg/day.  These associated trackets create a proximate quad.
  \item{} Apply the link testing algorithm described above to evaluate the
    minimum $\chi^2(s,w)$.  If $\chi^2_{min}$ is sufficiently small
    (nominally {\tt chimax} $<$25) continue consideration of this pair of
    tracklets.  The covariance matrices include both the projected
    uncertainties from Eq~\ref{eq:cov} added to the covariance from
    the linear fit to the extrapolation grid in $s,w$.
  \item{} Perform a \PUMA\ fit to all the detections from the pair of
    tracklets.  If the \PUMA\ reduced $\chi^2_\nu$ is sufficiently small
    (nominally {\tt chinmax} $<$25) record this pair of tracklets as a ``quad''.
    (A quad has 4 detections when \PUMALINK\ creates tracklets
    from internally linked pairs of detections.)
    Also compute the probability that the detections from this quad
    arise from one ($\chi_{\nu s}^2$) or two ($\chi_{\nu d}^2$) stationary
    objects.  (\PUMALINK\ treats stationary transients as moving
    objects with zero angular velocity, and $\chi_{\nu s}^2$ permits odds
    testing between stationary and moving hypotheses.)
  \item{} Once \PUMALINK\ has tested all pairs of tracklets and
    assembled all quads, \PUMALINK\ groups all quads according to
    whether they share a detection with some other quad in the group:
    different groups have no detection in common.  Within each
    detection group \PUMALINK\ examines the result of the \PUMA\ fit
    to each quad at the reference time, and groups all quads which are
    very close to one another (nominally 0.01~deg and 0.002~deg/day).
    Quads which are close but have inconsistent detections from a
    single epoch are not allowed to group, so a mis-linkages are not
    allowed to group with correct linkages.  The detections from
    singleton quads and groups that are subgroups of a larger group
    are dissolved into the bigger group.
  \item{} A final \PUMA\ fit using the detections from each
    group with more than 4 detections is recorded.
  \item{} \PUMALINK\ writes the results for each individual quad and
    the aggregate results from each quad group.  The detections that
    make up each quad or group are written as a comma separated list
    which may be used as a tracklet designation for a future run of
    \PUMALINK.  Note that the fundamental output from \PUMALINK\ is
    the file with individual quads because groups with more than 4
    detections can only exist for applications that have more than 2
    detections per epoch.
\end{itemize}

We desire three features of any linkage procedure: 1) high probability of
identifying correct linkages, 2) speed of execution, and 3) low number of
false linkages relative to good ones.

\PUMALINK\ prioritizes these three features in order.  For false
alarms, the \PUMA\ values of $\chi^2_\nu$ for the hypothesis of asteroid
orbit and the $\chi_{\nu s}^2$ and $\chi_{\nu d}^2$ values testing a mis-linkage of
one or two stationary objects provide a probability that a linking is
correct or not.  $\chi^2$ has a distribution of
$\frac{1}{2}\exp(-\frac{1}{2}\chi^2)$ for $N=8-6=2$ degrees of freedom
for a quad, so the nominal threshold of $\chi^2_\nu<25$ will pass many
false alarms into the output file that need to be rejected by the user.

The choice of parameter thresholds by the user may be used to trade
completeness against execution speed, and the best choice depends on
the input data.  \PUMALINK\ is most sensitive to two dimensionless
numbers from the data: $\Delta t / \delta t$, the ratio of
extrapolation interval to intra-tracklet interval, and $w\Delta t$,
the ratio of the extrapolation time and the notional ``collision
time''.  When $\Delta t / \delta t \gg 200$ it may be necessary to
widen the tolerances for the kd-tree first association of tracklets.
When $w\Delta t$ approaches 1 the linearity approximation degrades and
it may be necessary to increase the kd-tree search tolerances or else limit
the extrapolation grid away from very small values of $\rho$ or very
large values for $|\dot\rho|$ which make $w$ large.  Of course, \PUMALINK\ can be
run multiple times with different extrapolation grid ranges,
for example 0.1--4~AU and 0.01--0.1~AU.

Obviously the number of input detections and the limiting angular
velocity for forming tracklets affects the execution speed.  For many
datasets the execution time for \PUMALINK\ is dominated by the
tracklet extrapolation stage, which takes approximately 150
microseconds per tracklet.  The number of tracklets goes as the square
of the product of the number of detections, the limiting angular
velocity, and the typical intra-tracklet interval.  For typical sky
surveys the number of tracklets is roughly in the range $N_{track}\sim
10^{-5}\,N_{det}^2$, so the \PUMALINK\ execution time is roughly
1--100 nanosec per $N_{det}^2$.

\subsection{Pumalink performance}

We performed a number of tests of \PUMALINK\ on a variety of datasets
that illustrate its performance.  The first test is a set of
normal ATLAS operations from MJD 58995 to 59005 (new to full moon),
the second test is a set of special ATLAS observations with 110~sec
exposures of a particular field near opposition from MJD 59991 to
60003 (last to first quarter moon), and the last test was an LSST
simulation kindly provided by Ari Heinze and Mario Juric spanning
future MJD 60601 to 60616 (full to new moon).

ATLAS carefully notes the presence of all known asteroids
(``kast'') for each exposure, and the completeness of known asteroids
is very high at the ATLAS limiting magnitude, so the kasts allow us to
evaluate object detection probability and false alarm rates.

\subsubsection{ATLAS nights}

From MJD 58995 to 59005 the two Hawaii observatories stepped between
four declination bands, Maunaloa and Haleakala following each other so
each band was observed every other day.  Each field was nominally
targeted 4 times on each night, and the time intervals $dt$ between
successive observations were approximately 5, 15, and 30~min.  We ran
\PUMALINK\ on each night against every other night that covered the
same Dec band, so the lag times spanned by the two \PUMALINK\ epochs
were 2, 4, 6, 8, and 10 days.  We made no attempt to condense the 4
visits to each field every night directly, and simply let
\PUMALINK\ deal with all possible pairs of detections for each object.
ATLAS detections are each tested by up to five different functions
depending on SNR: fixed PSF, adjustable PSF, trailed PSF, long streak
PSF, and negative flux, and then a preliminary classification for each
detection assigns it to one of 8 possibilities, of which one is
corresponds to a real object that is not a variable star.
Detections were excluded if this basic ATLAS classifier probability
of being real was lower than 10\%, or if they were output from the
routines that fit trailed detections.  

The 21 declination band visits created 45 overlapping pairs of nights,
the median number of detections given to \PUMALINK\ for a pair of
nights was 2.5M, the
median number of tracklets formed was 10M, and the median execution
time was 1800 sec (1 core, 3GHz).  We deem a kast to be
``linkable'' by \PUMALINK\ when at least two detections exist for
it from each epoch so that two tracklets may be formed.
Among the pairs of nights, the median fraction of kast actually found by
\PUMALINK\ was 98.7\%.  This success fraction drops quadratically as a
function of lag time between nights, changing from 99.7\% at a lag of
2 days to 94.5\% at a lag of 8 days.  Overall, of the 32859 kasts that
were linkable between a pair of nights, \PUMALINK\ found 32585,
99.2\%.

To quantify the performance we use 
a binary confusion matrix
where {\tt TP} and {\tt TN} designate correct classifications of
actual positive and negative cases as positive and negative, and {\tt
  FN} and {\tt FP} are erroneous classifications of actual positive
and negative objects as negative or positive.  A {\tt FN} (``miss'')
counts against the detection probability and a {\tt FP} adds to the
false alarm rate.

We calculate a ``probability of detection'' (PD, also known as
``recall'') as the fraction of all linkable kasts that appear in at
least one quad, i.e. {\tt TP}/({\tt TP}+{\tt FN}).  The number of {\tt
  TN} is ill-defined and potentially vast, so we define our ``false
alarm rate'' (FAR) as the fraction of all quads which do not match any
kast or which mis-link different kasts, i.e.
{\tt FP}/({\tt TP}+{\tt FP}).  This is also known as (1$-$``precision'').

\begin{figure}[ht]
\begin{center}$
\begin{array}{ccc}
\includegraphics[width=2.3in]{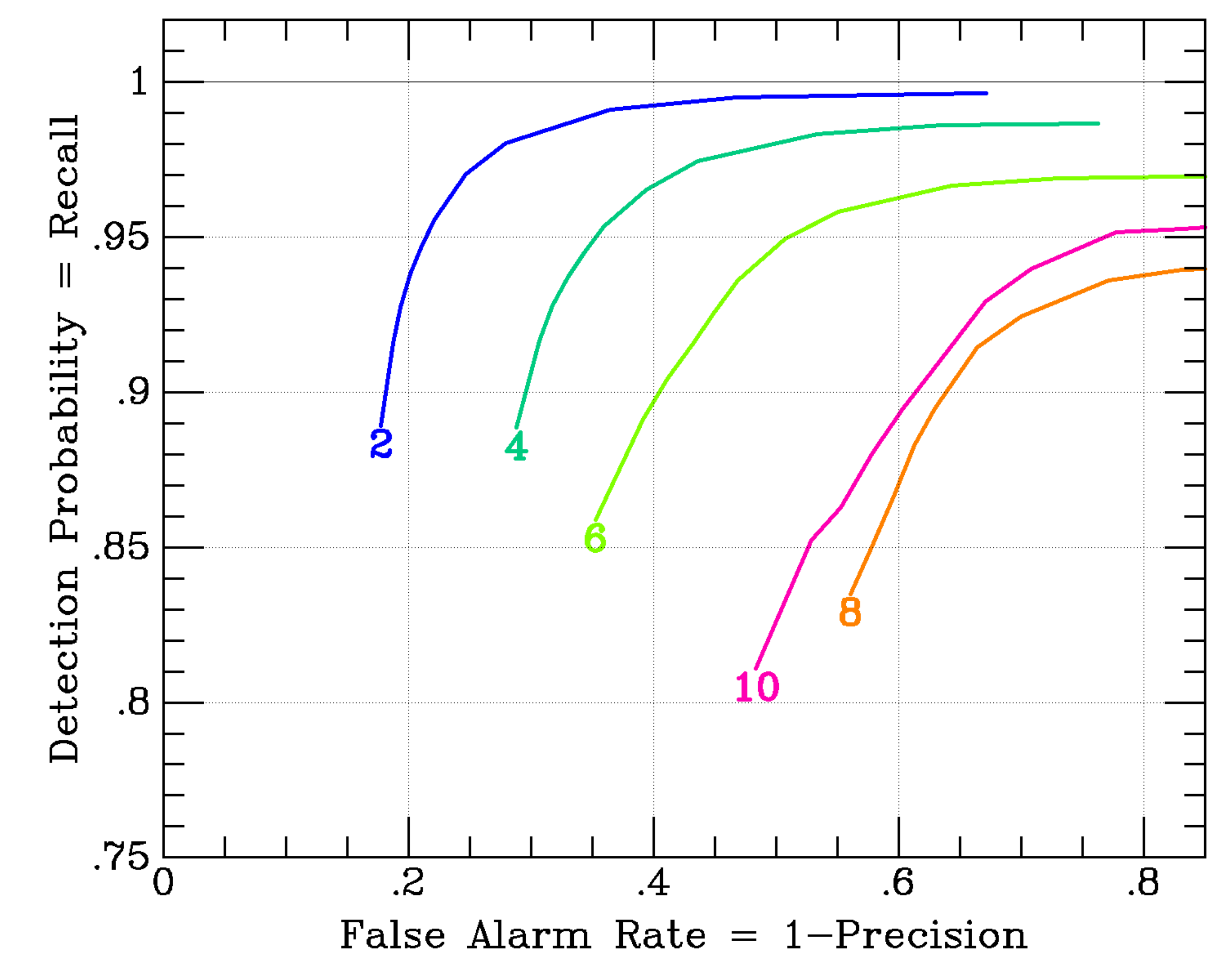}
\includegraphics[width=2.3in]{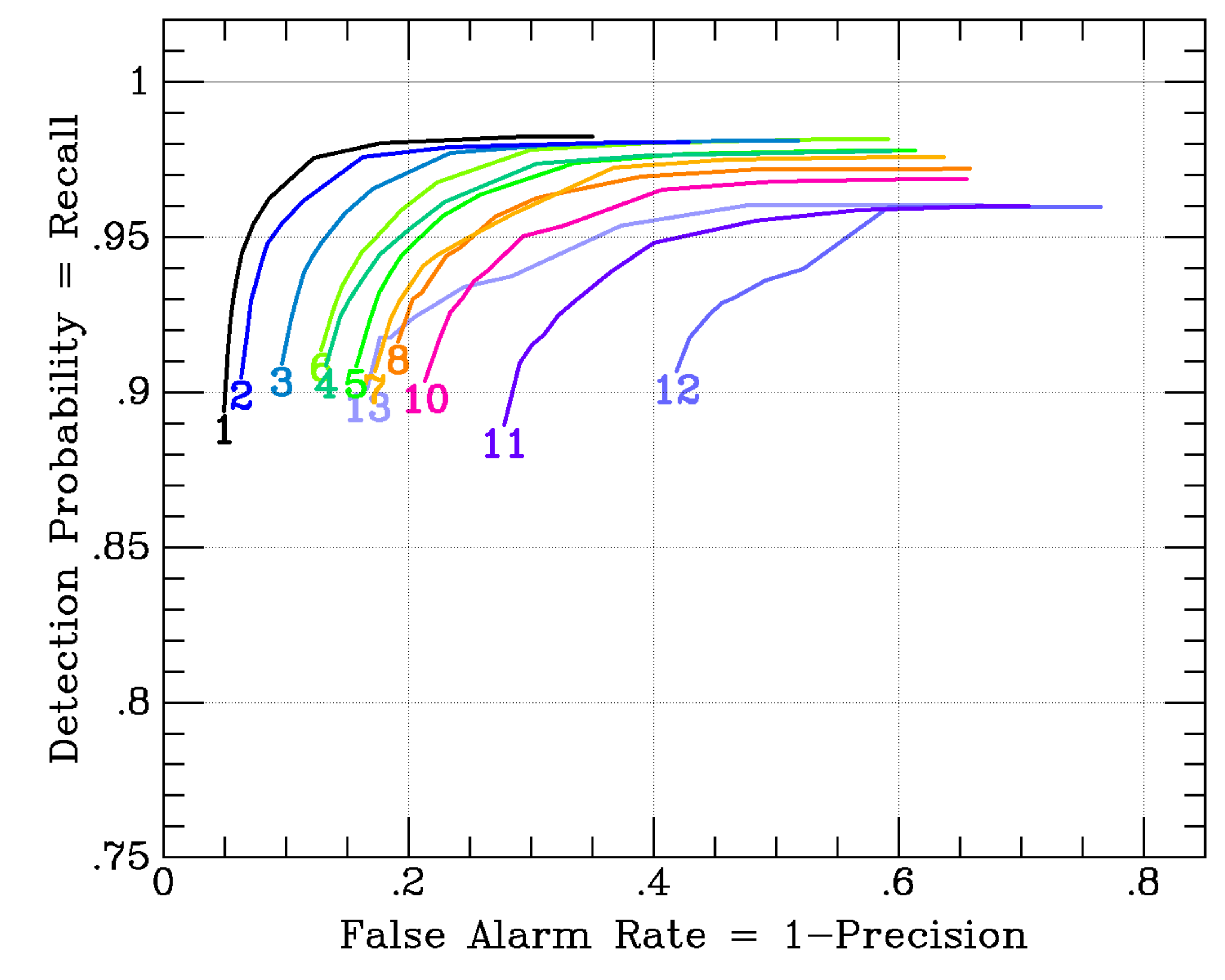}
\includegraphics[width=2.3in]{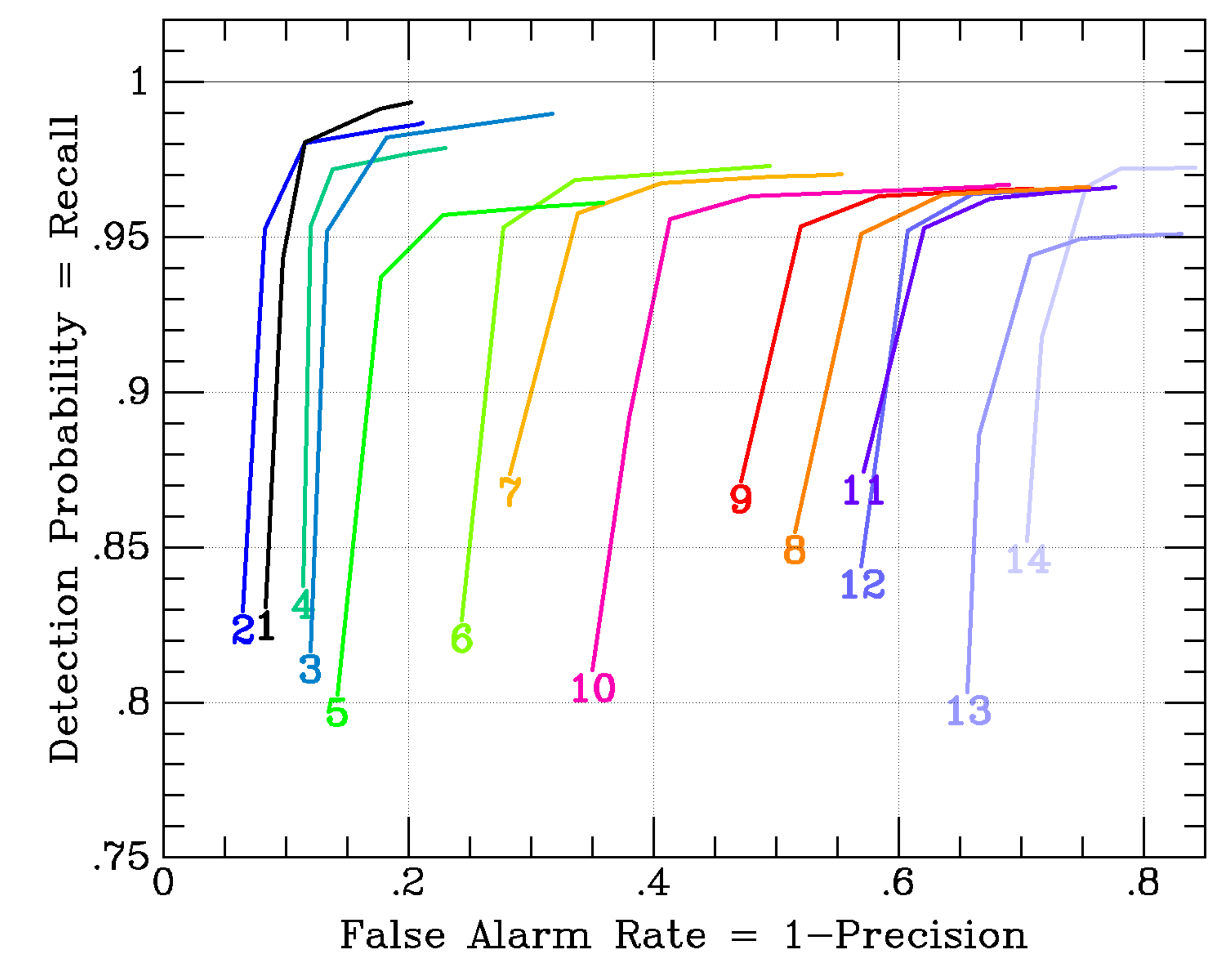}
\end{array}$
\end{center}
\caption{This figure shows the detection probability versus false
  alarm rate (``receiver operating characteristic'' or ROC curve) for
  the performance of \PUMALINK\ on the three datasets as a function of
  the {\tt chinmax} $\chi^2_\nu$ threshold parameter.  The left panel
  shows the 11 full ATLAS nights, the middle panel the ATLAS
  opposition field experiment, and the right panel the LSST simulation.  For
  these ``detection probability'' is the fraction of all linkable
  kasts that are linked by at least one quad and the ``false alarm
  rate'' is the fraction of all quads that do not associate
  with any kast.  The colors and labels show the inter-epoch lag
  time: black through magenta are lags from 1 to 10 days, then lighter
  shades of purple show lags to 14 days for the opposition and LSST
  tests.}
\label{fig:roc}
\end{figure}

As noted above, the performance depends on the lag time $\sim2\Delta t$
between pairs of nights for the nominal parameters.  Of course the
PD and FAR depend on many different input parameters, but it is
illustrative to vary only the \PUMA\ $\chi^2_\nu$ threshold for keeping a
linkage and generate the set of curves of PD and FAR illustrated
in Figure~\ref{fig:roc}.

This experiment has $\Delta t/\delta t$ as
large as $\sim1500$, which causes the PD and FAR performance to
degrade for the biggest lags.  Opening up the tolerance for the
kd-tree linkage could recover some of this performance at the cost of
more compute time and FAR.

Although the PD reaches very respectable values, the FAR is never very
small for this full night test.  An extrapolation by $\Delta t/\delta
t\sim100$ or larger will always admit false linkages that are
statistically reasonable.  However, a 50\% false alarm rate is
actually very manageable.  Each candidate linkage typically is
subjected to additional image scrutiny (human or machine), and there
are other statistics that can be considered, even from
\PUMALINK\ (such as $\chi_s^2$ and $\chi_d^2$, or using a process of
elimination on detections to reject mis-linked quads).  Removal of
false alarms (potentially at the cost of real detections) depends on
the application and is beyond the scope of what \PUMALINK\ can do by
itself, but we discuss FAR reductions strategies in the next section.

Figure~\ref{fig:roc}
treats each pair of epochs separately, and combining the linkages from
multiple pairs of nights does improve the PD and suppress the FAR.
In addition, given more than two epochs \PUMALINK\ can be run
hierarchically, taking advantage of the performance at small lag time
before running \PUMALINK\ over a bigger lag time using the output
linkages from a previous iteration.

The distribution of the magnitudes of the linked kasts peaks at
$V\sim18.6$ and the 50\% detection rate occurs at $V\sim19.0$.  The
nearest kast found by \PUMALINK\ with the default parameters during
this exercise was at 0.30~AU and the fastest was 2.7~deg/day.
NEO~513170 was also present in these observations at 0.11~AU and
4.3~deg/day.  Although this NEO was missed by the nominal values for
the kd-tree search of 0.2~deg and 0.5~deg/day, it was found when these
were increased to 0.3~deg and 1.0~deg/day.  Increasing these limits
also increased the execution time by about 50\%.

\subsubsection{Opposition experiment}

Between MJD 59990 to 60003 ATLAS observed a particular field every
night fairly close to opposition so that there were many asteroids near
full illumination.  In addition, the exposure time was increased from
30~sec to 110~sec which increased the sensitivity by nearly a
magnitude.  On any given night there were 4--8 observations of the
field, sometimes from different observatories.  The time interval
between each observation was kept uniform at $\sim$30~min.

These observations over 14 days created 91 pairs of nights to be
compared, the median number of detections given to \PUMALINK\ from a
pair of nights was 20k, the median number of trackets formed was 300k,
and the median execution time was 35 sec.  The median fraction of kast
found by \PUMALINK\ was 98.3\% and the average was 97.4\% but it
varies less with lag time compared to the full night experiment
because $\Delta t/\delta t$ is smaller, never exceeding $\sim 300$,
thanks to $\delta t$ never smaller than 30~minutes.

Of the 1295 kasts which were linkable between at least two nights,
\PUMALINK\ found 1263 with the default parameters.  The missing kasts
mostly were ones that had a single detection from South Africa and then
another from Chile 4 hours later, and that exceeds the default time
difference of 0.1~day to form a pair.  If that time interval between
two detections to form a test tracklet is increased, \PUMALINK\ finds
1292 of 1295 kasts, but at the cost of a higher false alarm rate.

The ROC curve seen in the middle panel of Fig~\ref{fig:roc} provides
more insight into the differences between the opposition test and the
full night test.  The FAR at a given PD is considerably smaller for
the opposition fields than for the full nights.
This is the result of 
$\Delta t/\delta t\lesssim 300$: it is much less common for a false
alarm linkage to be statistically consistent with the observations
when the extrapolation is less extreme.

The distribution of the magnitudes of the linked kasts in these
opposition fields peaks at $V\sim19.6$ as a result of the increase in
exposure time, and the 50\% detection probability occurs at
$V\sim20.0$.  The nearest and fastest asteroid in this small
opposition field was 2008CX189 (0.75~AU and 0.57~deg/day); 2008CX189
was correctly linked by \PUMALINK\ with the nominal parameters.

\subsubsection{LSST simulation}

The Rubin Telescope\footnote{\tt https://rubinobservatory.org} is
scheduled to begin the Legacy Survey of Space and Time (LSST) sometime
in 2025\footnote{\tt ls.st/dates}.  \cite{2017ASPC..512..279J}
describe the LSST data management system, \cite{2021DPS....5310106J}
presents 10 year simulations of solar system moving objects
appearances in the LSST data, and \cite{2022DPS....5450404H} discusses
the application of HelioLinC to find these moving objects in the real
data.  Among the many science goals that will be addressed by LSST is
the detection of large numbers of asteroids, particularly NEOs.
Because the schedule will be driven by many competing projects, the
observation cadence for asteroid detection will be challenging.
Nevertheless there is no doubt that the LSST will be extremely
powerful, and software to analyze simulated LSST data is well
advanced.

Ari Heinze and Mario Juric generously provided us with one such
simulated dataset that spans future MJD 60601 to 60616.  This differs
from the ATLAS data already considered because there are no false
alarm detections, just 5.3M detections of 995082 different asteroids
with simulated astrometric error and magnitudes.  Therefore
``false alarm'' might better be termed ``mis-linkage'' for this
test.  These detections are
all labeled by an object ID (which we ignore until it is time to
verify performance).  We organized these into distinct observations
(same exposure start time) and nights, and create detection files in
{\tt TRD9} format.  We assigned an astrometric error of 1~arcsec
multiplied by the simulation magnitude uncertainty with 0.015 mag added
in quadrature (e.g. $\sim0.03$~arcsec at $m\sim22$).

These 15 simulated nights make 104 pairs of epochs on which we ran
\PUMALINK\ with the nominal parameters.  The median number of
detections from the pairs of nights was 730k, the median number of
tracklets formed from an epoch pair was 8.3M, and the median execution time
for a pair of nights was 1000~sec.  Of the nearly million asteroids,
only 427,827 appear at least twice on two different nights and are
therefore linkable by \PUMALINK.  Of these, the total number linked
between at least two nights was 420,708, for a gross PD of 98.3\%.

\PUMALINK\ faced a number of challenges with the LSST data.  Some of
the $\delta t$ intervals were extremely brief, and the extrapolation
ratio of $\Delta t /\delta t$ exceeded 12,000 for some tracklets.
LSST also spends time on a number of nights observing the ``deep
drilling fields'' which have as many as 96 exposures at the same
location in a night.  Our naive application of \PUMALINK\ creates
96-choose-2 possible tracklet pairs from such a night for an asteroid
in that field, and between two such nights each asteroid creates 20M
different pairs of tracklets which require \PUMA\ evaluations.
Operating on such pairs of nights \PUMALINK\ can use more than 100G of
memory and 200,000 seconds to complete (although it eventually does
succeed).  Evidently \PUMALINK\ could be used more efficiently with a
more judicious selection of input.

The right panel of Figure~\ref{fig:roc} shows more detail on the
\PUMALINK\ performance.  As
with the other two experiments the PD at a given FAR and the FAR at a
given PD degrade for bigger extrapolation ratios.  When \PUMALINK\ is
run hierarchically, using the linkages from a pair of nights as input
tracklets such as 60103--60105 and 60109--60113, the false alarm
rate is zero.

The nearest and fastest linkable asteroid in the LSST simulation had
ID S0000ssOa at 0.022~AU and 18.5~deg/day on MJD 60601.  It is
instructive to examine how \PUMALINK\ deals with it.  In the LSST
simulation it also appears more than once per night on 60603 at
0.042~AU and 5.0~deg/day, on 60605 at 0.061~AU and 2.7~deg/day, on
60606 at 0.073~AU and 1.6~deg/day, on 60610 at 0.115~AU and
0.7~deg/day, and on 60611 at 0.126~AU and 0.6~deg/day.
Recall that as we are running it here \PUMALINK\ first must form a
tracklet on a day within an angular velocity limit
and then try to link two different days.

\PUMALINK\ cannot form a tracklet on 60601 and 60603 with the default
angular velocity of 5~deg/day, but increasing it to 6~deg/day does
form a tracklet on 60603.  In order for \PUMALINK\ to link 60603 and
60605 it must be given kd-tree search limits of 1~deg and 4~deg/day
instead of the defaults of 0.2~deg and 0.5~deg/day.  This increased
search area is necessitated by S0000ssOa's deceleration from 5~deg/day
to 1.6~deg/day over that two day time span and the limitations of
\PUMA's short arc extrapolation.  Between 60605 and 60606
when S0000ssOa moves from 0.06 to 0.07~AU \PUMALINK\ needs a slightly
increased search area of 0.3~deg and 0.5~deg/day.  Between 60610 and
60611 \PUMALINK\ finds S0000ssOa and links it correctly.

Once the detections for S0000ssOa are given to \PUMA\ for testing, the
result has $\chi^2_\nu{<}1$ on all pairs of nights.  Of course, S0000ssOa
would be trailed on the first three nights, and passing a greatly
reduced set of trailed detections to \PUMALINK\ would take negligible
processing time.

The only other asteroid in the LSST set closer than 0.1~AU and moving
more slowly than 5~deg/day was S0000EOMa at 0.081~AU moving at
4.41~deg/day, and it was linked by \PUMALINK\ with the default parameters.

\subsection{Optimizing \PUMALINK\ and coping with false alarms}

\subsubsection{Parameter optimization}

\PUMALINK\ can be optimized for each application by adjusting its
operation parameters from the defaults.  As always the optimization
needs to balance the goals of high PD, low execution time, and low
FAR.  The three examples above illustrate how \PUMALINK\ performs for
three representative test cases with the default parameters.  This
optimization is application specific because \PUMALINK\ is not witting
of many survey characteristics, for example how many times a given
field was observed on a given epoch or what fraction of the detections
given to \PUMALINK\ are expected to be false.  This subsection will
briefly describe the operating parameters and how each is likely to
affect the three goals.

Given the \PUMALINK\ results on several representative experimental
datasets, we return to the discussion of Sec~\ref{sec:sw} illustrated
in Fig~\ref{fig:plflow} and offer guidance on the different
parameter's function.

\begin{itemize}
  \item{} Pair formation from all the detections is governed by the
    parameters {\tt dtmax} and {\tt omega}.  We saw a case with the
    opposition experiment where {\tt dtmax} had to be increased to
    support a 6 hour time interval for a tracklet, and a case with the
    LSST experiment where {\tt omega} had to be increased to 6~deg/day
    to find a nearby asteroid.  The execution time goes goes
    approximately as the square of the product of {\tt dtmax} and {\tt
      omega}, as does the false alarm rate.  For the case of very fast
    moving, trailed detections it is possible to set {\tt omega} to a
    very large value and simply test all possible pairs with
    \PUMALINK, at a cost of about $\sim\hbox{10 nsec}\; N_{det}^2$.

  \item{} The accuracy of the linear fit to \PUMA\ extrapolations of
    tracklets is affected by the choice of the extrapolation grid.
    Since the predicted locations and error grows with $s=1/\rho$, it
    is more efficient to use a minimal grid that does not extend to
    very small distance, but this runs the risk of not
    linking a very nearby object.  When \PUMALINK\ is given a
    user-defined set of tracklets there may not be much latitude to
    explore a large grid, and many of the grid points may be rejected
    by the \PUMA\ extrapolation because they exceed the {\tt chieph}
    parameter.

  \item{} \PUMALINK\ sorts the extrapolated positions using a kd-tree
    in order to judge which tracklets should be tested with the
    pumalink $\chi^2$ calculation of Eq~\ref{eq:chi}.  We have not yet
    found a way to exploit the very tiny volume that these
    extrapolations occupy in 6D space, and instead use an enclosing
    volume described by the parameters {\tt dxmax} and {\tt dwmax} for
    the allowable difference in unit vector and tangential velocity.
    These default parameters are not adjusted by \PUMALINK\ for the
    time interval $\Delta t$, which is the main reason for the
    degradation in PD with increasing $\Delta t$.

    For example, by doubling {\tt dxmax} and {\tt dwmax} for the ATLAS
    full night experiment, the median PD changed from 98.7\% to
    99.8\%, and the parabolic decline of PD with with lag time
    improved from 99.7\%--94.5\% for lags of 2--8 days to
    99.9\%--99.3\%.  This widened acceptance also caused the execution
    time to become 5 times longer and increased false alarms.

    The choice of kd-tree search parameters is the primary tradeoff
    between performance closer than 0.1~AU and processing time, and
    the \PUMALINK\ code is not as efficient as it could be in this
    regime.  With the default parameters \PUMALINK\ spends
    $\sim$50~microsec per kd-tree search to find each close pair of
    tracklets that is then passed to the pumalink algorithm, but only
    $\sim$0.25~microsec for the pumalink calculation to accept or
    reject them.  More efficient sorting and pairing would allow the
    kd-tree parameters to be relaxed without a performance penalty.

  \item{} Once a pair of tracklets is known to be close in 6D space at
    the reference time, the pumalink $\chi^2$ comparison tells us
    whether there really is a $\rho,\dot\rho$ for which the two
    tracklets are statisically close to one another.  The parameter
    {\tt chimax} governs whether the pair will be rejected or kept as
    a possible quad.  {\tt chimax} may be increased to avoid losing
    very nearby objects whose puma extrapolation may not be well
    matched by a linear function of $s,w$, but at the cost of more
    false alarms.

  \item{} Quads that have been passed by the pumalink calculation then
    have their detections fitted by \PUMA, and they are kept if
    $\chi^2_\nu$ is less than {\tt chinmax}.  Again the false alarm rate
    rises rapidly with statistically unlikely values for
    {\tt chinmax}, even though exceptional real objects may require it.

  \item{} The detection grouping carried out by \PUMALINK\ is
    straightforward, but the grouping of quads by interpolation to the
    reference time is governed by the parameter {\tt grptol}.  It is
    possible to widen this, at the cost of putting multiple objects in
    the same group, or to tighten it, at the cost of failing to groups
    multiple quads of the same group.
\end{itemize}

\subsubsection{Balancing detection probability against false alarm rate}

After \PUMALINK\ completes, there will typically be many false alarms.
It is then the user's responsibility to decide how to set the
acceptance criteria: reducing false linkages almost certainly will
also remove linkages of real objects.

It may be possible to adjust one or more parameters as in the ROC
curves of Fig~\ref{fig:roc} to optimize PD and FAR, but the optimum is
depends on the application.  In some cases a detection might be so
valuable that a huge FAR is acceptable and it's worth spending
additional resources to sort out the false alarms.  In other cases
even a small FAR not be acceptable because of harm to reputation.  In
the minor planet community, concern over wasted followup observations
spent chasing false alarms causes surveys to be very conservative
about reporting possible objects to the MPC (which unfortunately also
hinders the possibility of advancing two uncertain linkages to a
certain one).

The detections that are provided to \PUMALINK\ typically sample both
real and false populations.  The real population has a distribution of
object brightness that typically grows as some sort of power law, as a
result of distance and from intrinsic luminosity distributions.  The
false population results from statistical fluctuations, both Gaussian
noise and non-Gaussian effects in difference images.  Therefore the
number of false versus real detections is very different
as a function of the ``SNR limit'' cutoff, with the ratio of
false to real detections growing rapidly for fainter cutoffs.  This
cutoff choice is outside of the purview of \PUMALINK; nevertheless
\PUMALINK\ will form many false linkages from false detections.  As
long as the execution time is not too burdensome this can be
advantageous, however, because a quad can now be evaluated on the
basis of the probability that all four detections are real, not just
each one separately.

To get a sense of the numbers, a given ATLAS night has $\sim10^{11}$
potential detections, corresponding to every positive-going excursion
on a difference image.  This is reduced by a factor of $\sim10^{5}$ to
$\sim10^{6}$ by the detection process, which tries to identify every
positive-going excursion that is consistent with the PSF shape,
subject to an SNR cutoff that balances retention of real objects
versus admitting excessive false alarms.  In fact the ATLAS software
detects and keeps all PSF-like excursions to the 3-sigma level, but
only passes ones that exceed $\sim5$-sigma.  For Gaussian statistics
about one in three million events exceed 5-sigma, so this reduction of the
number of events by only $10^5$ illustrates the heavy tails of the
actual distribution of events in the ATLAS images.

All ATLAS detections are accompanied by measurements of the SNR of
the detection, the extent of the detection if it is trailed, and the
$\chi^2_\nu$ agreement between the detection and the local PSF shape.
Additional information available for each detection includes the image
mask (including the location of cross-talk artifacts and saturated
pixels), the image before differencing, the ``wallpaper'' image from
previous observations, and the list of all known stars and asteroids.
All of this information is given to a classifier that assigns
probabilities that the detection may be real or false, and if false
what it might be.

The ATLAS classifier does a decent job of distinguishing transients and
variable stars from image artifacts, but it is not perfect and using it to
cull the input detections given to \PUMALINK\ has a cost.  Our
experiment with \PUMALINK\ on the ATLAS nights used a classifier
threshold of only 10\% that the detection is real, and in so doing
lost 12\% of real, linkable asteroids that \PUMALINK\ might have
captured.  This tradeoff exists so that MOPS can complete on a short
timescale.
The full night ATLAS experiment also removed all detection information
indicating that an object trailed during the exposure (consistent with
the {\tt omega} cutoff).  These cuts remove only a small fraction of
all detections and do not significantly aid \PUMALINK, and therefore
should be carefully reconsidered when optimizing \PUMALINK\ for the
ATLAS application.

Given these detections, \PUMALINK\ forms tracklets within {\tt
  dtmax} and {\tt omega} at each epoch, producing $\sim10^7$ for
each epochs.  This is a straightforward proximity search that
makes no attempt to decide when a pair is plausible.  For example the
fact that detections might have a similar brightness or slight
extendedness in the same direction is ignored.

There are altogether $\sim10^{14}$ pairs of tracklets which could form
a quad, and the algorithm described here manages to distill this down
to $\sim10^5$ output quads, depending on sorting and efficient
algorithms to kept the execution time short.  As we have seen, within
this set of quads the probability of keeping real objects is extremely
high, nearly 100\%, particularly for few day time lags or widened
acceptance criteria.  However, among these quads there are typically
only $\sim10^4$ real objects and so $\sim90$\% of quads and groups in
the \PUMALINK\ output are linkages of false detections or mis-linkages
of real objects.  There are a few quads which can be rejected with
high certainty because \PUMALINK\ finds them to be consistent with
stationary objects ($\chi^2_{\nu s}<10$), but for the most part reducing
the false alarm rate involves moving down the ROC curve and some real
objects will be lost.

The simplest method to reduce the false alarms is to reject linkages
with large \PUMA\ $\chi^2_\nu$, i.e. move along the ROC curve of
Fig~\ref{fig:roc}.  For the ATLAS nights we have seen that the knee in
the ROC curve occurs at false alarm rates of around 50\% and the loss
of several percentage points in detection probability.  We can do substantially
better, however, by
\begin{enumerate}
  \item{} formulating a better probability for each detection using
    all the information available,
  \item{} making decisions based on \PUMALINK's groups,
  \item{} re-examining the images themselves and re-classifying detections, and
  \item{} seeking more detections that join onto \PUMALINK's linkages.
\end{enumerate}
We will use the ATLAS full night experiment to illustrate each of
these methods.

For ATLAS we have a number of pieces of
information about each detection that were not fully exploited in the
culling of input detections.  First and foremost are the SNR of the
detection and the $\chi^2_\nu$ result when the detection was tested
against the local PSF in the image.  Additional information comes from
the proximity to the edge of the detector where the noise enhancement
from flatfielding is not perfectly modeled as well as the classifier
output for faint detections.  From the distribution of these detection
characteristics for all the kast (which we know to be real) and
non-kast (which are mostly false) we can form an ad-hoc probability
function for detections (which is specific for ATLAS).

\begin{figure}[ht]
\begin{center}$
\begin{array}{cc}
\includegraphics[width=2.3in]{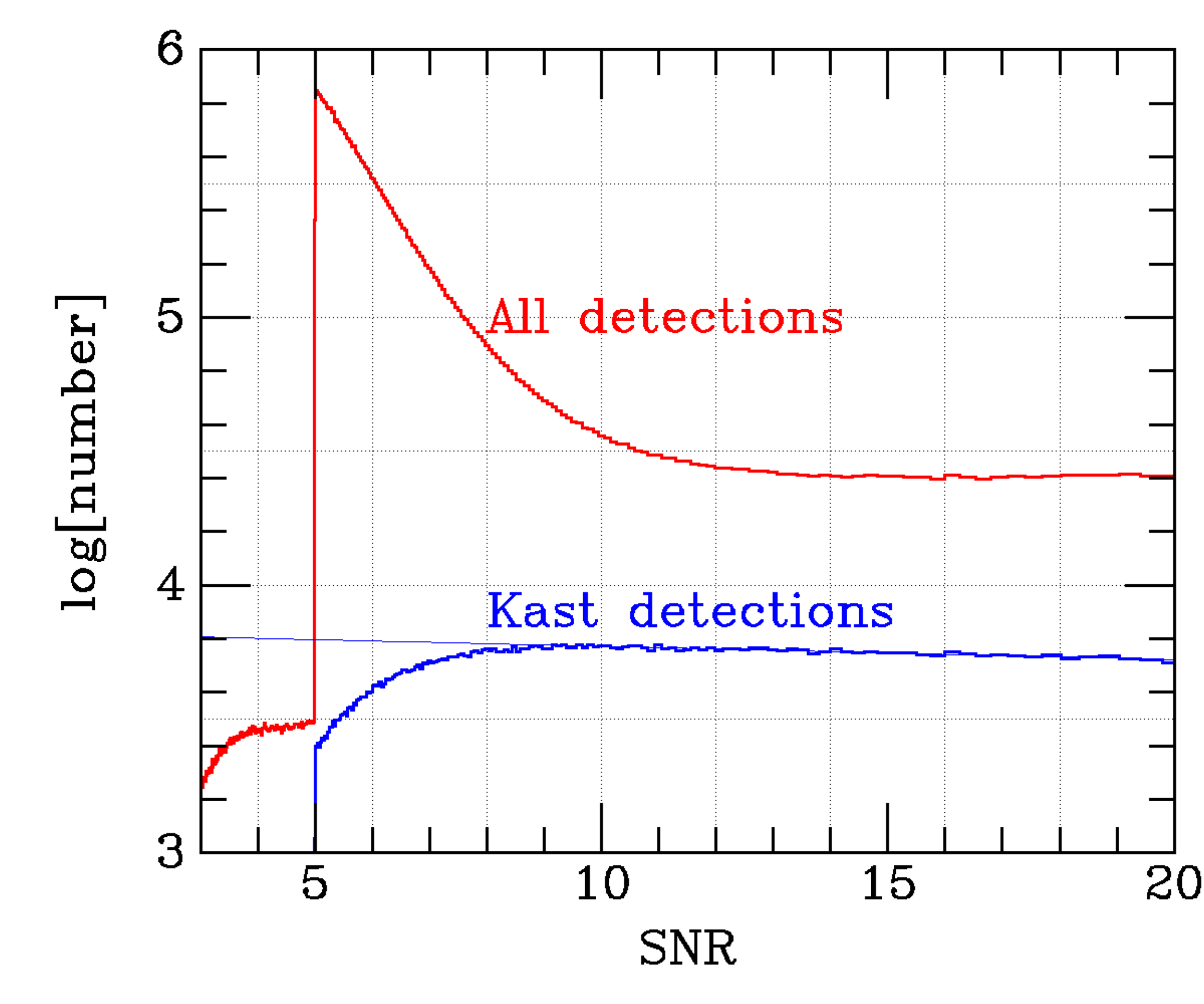}
\includegraphics[width=2.3in]{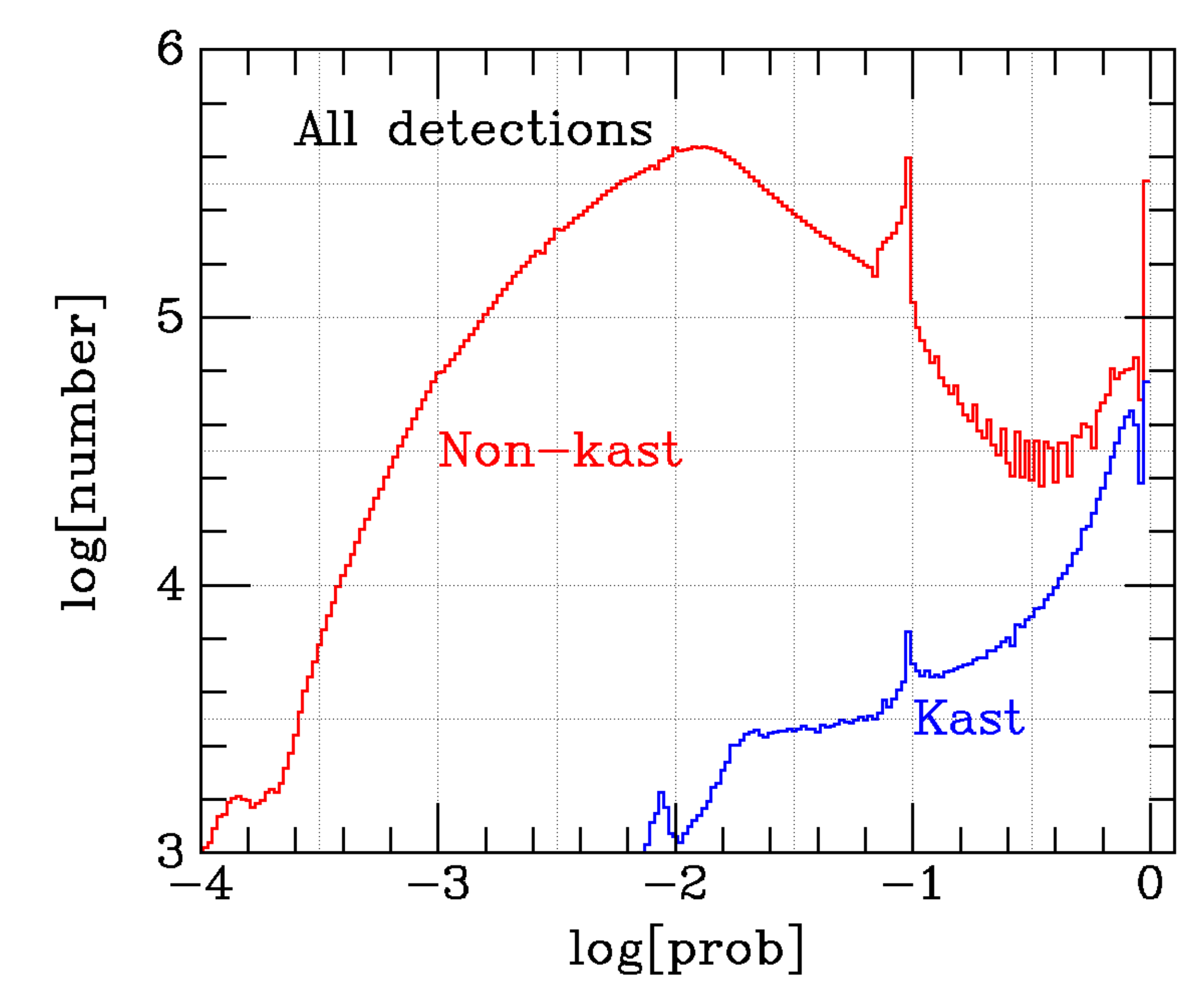}
\includegraphics[width=2.3in]{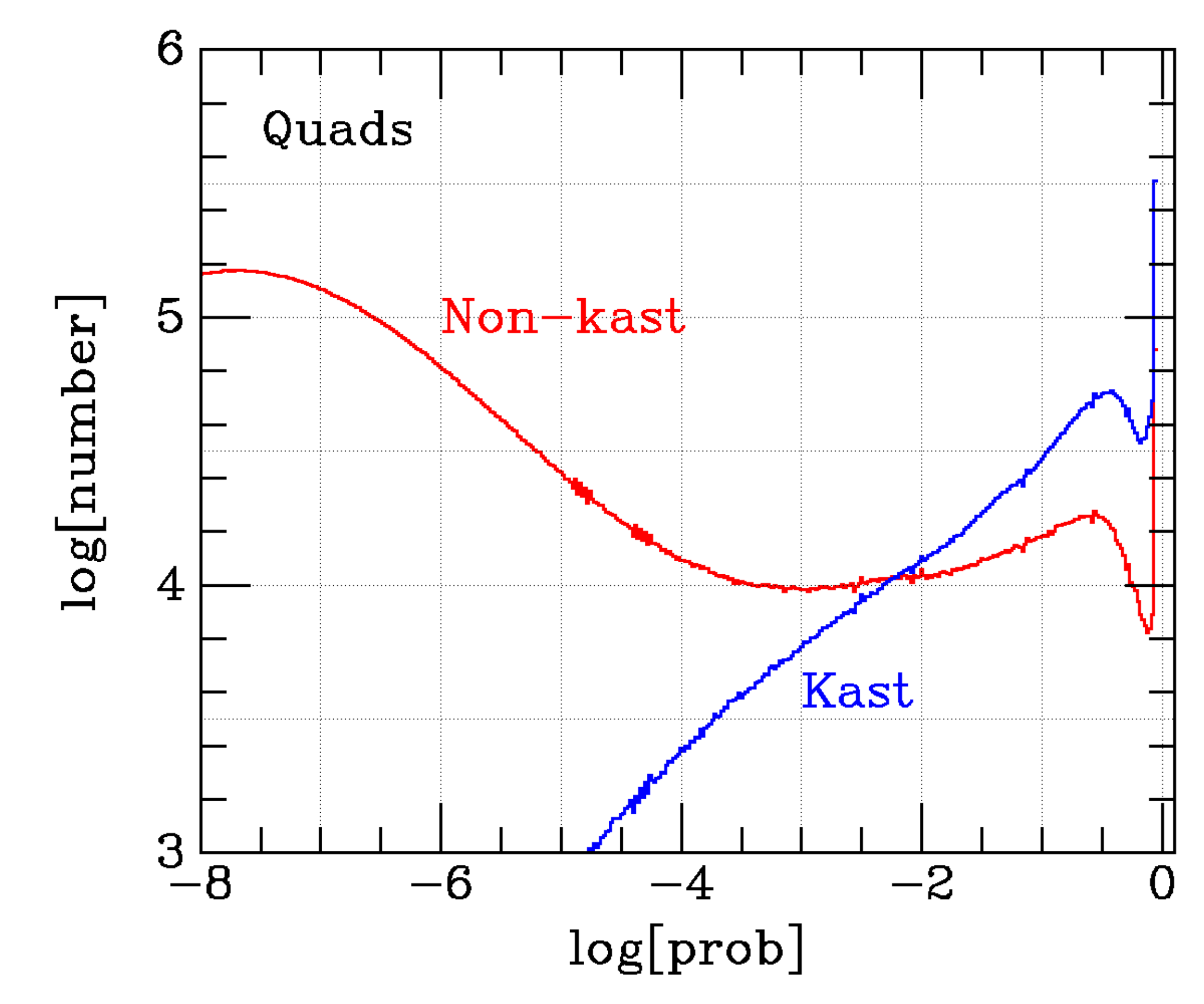}
\end{array}$
\end{center}
\caption{This figure shows the distribution of signal to noise for all
  detections (left), the distribution of the ad hoc probability
  function for all detections (center), and the aggregate
  probability function distribution for quads formed from the product
  of the probabilities of the 4 detections (right).  In the left panel the SNR
  distribution cuts off abruptly at 5 because of the threshold applied
  to input detections.  The count of kast detections rolls off below
  SNR~$\sim8$, and is down by about a factor of 2 at SNR~$\sim5$
  relative to the indicated extrapolation.  In the middle panel the
  little spike at $-1$ on the curves is caused by part of the ad-hoc
  probability function that puts a floor of 0.1 on the probability for
  detections with very large $\chi^2_\nu$ for the PSF fit (in order not
  to eliminate comets, for example).  In the right panel, the
  probability product is such a powerful discriminator that there are
  more kast quads at large probability than non-kast, unlike the
  distributions for each detection in the middle panel.}
\label{fig:snr}
\end{figure}

Figure~\ref{fig:snr} shows the distribution of the calculated SNR of
all the ATLAS detections, and the distribution of this ad-hoc
probability function for individual detections as well as the product
for the detections that make up possible quads.

\begin{figure}[ht]
\begin{center}
\includegraphics[width=2.3in]{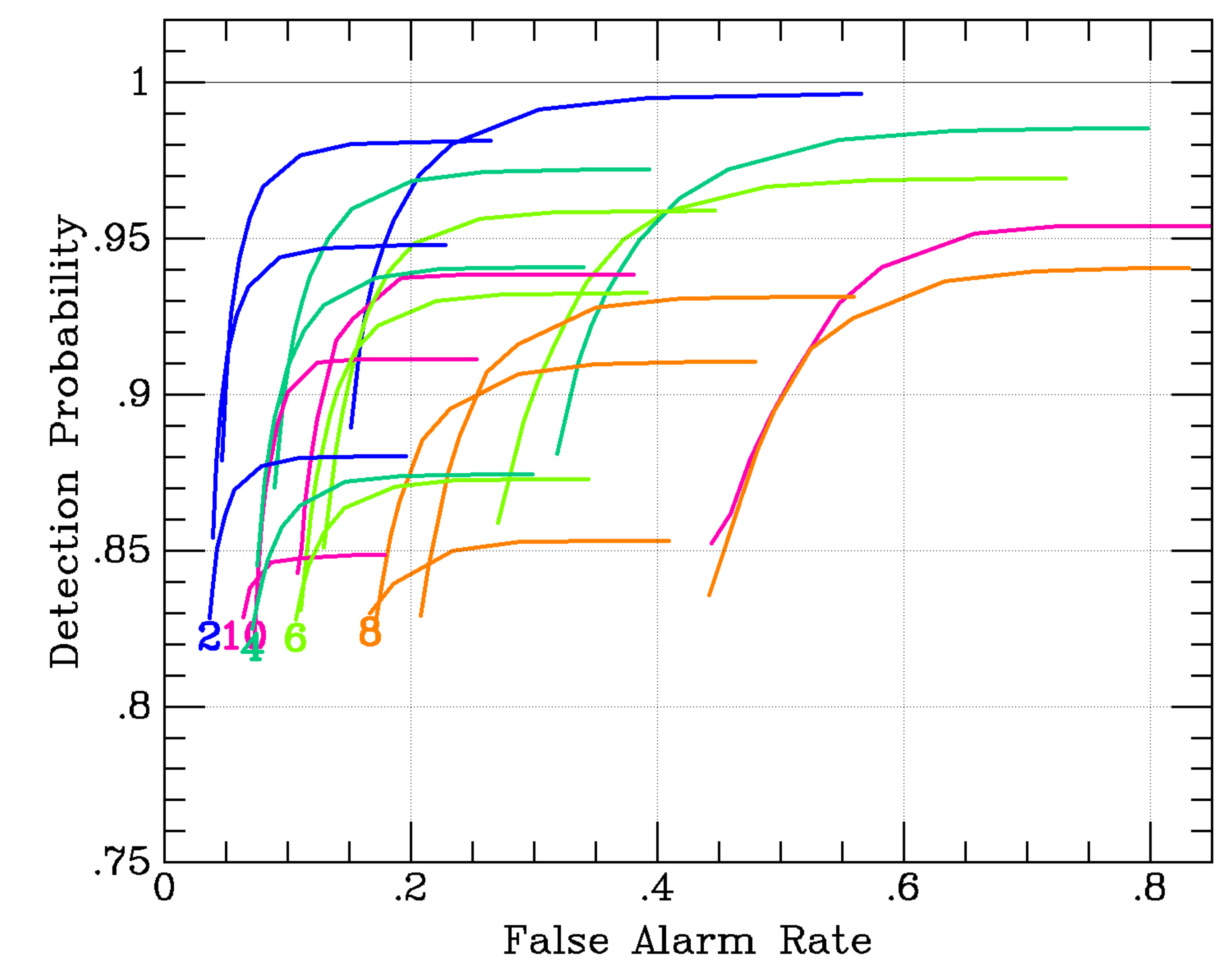}
\end{center}
\caption{This figure shows \PUMA\ $\chi^2_\nu$ ROC curves for all the
  quads from the full night ATLAS experiments.  As in
  Fig~\ref{fig:roc} the colors show the inter-epoch lag time: blue
  through magenta are lags from 2 to 10 days.  The rightmost set of
  curves in each color show the results without
  post-\PUMALINK\ probability cuts but with a cut on $\chi^2_{\nu s}$ to
  reject stationary objects and a cut to reject space velocity greater
  than 50~km/s.  Moving to the left and lower false alarm rate are
  curves restricted to quads with $\log(P)>-5$, $\log(P)>-4$, and
  $\log(P)>-3$.}
\label{fig:pdchop}
\end{figure}

We calculate the probabilities for each detection, multiply them
together for each output quad from \PUMALINK\ to create an aggregate
$\log(P)$, and then use different trial thresholds on that to select
``good'' quads.  For these ``good'' quads we calculate ROC curves
by varying the \PUMA\ $\chi^2_\nu$ cutoff, and make a new family
of ROC curves shown in Fig~\ref{fig:pdchop}.
As we saw in Fig~\ref{fig:snr} the false alarms are greatly suppressed
by requiring the detection probability product $\log(P)$ exceed some
threshold, and additional cuts from the \PUMA\ $\chi^2_\nu$ can reduce the
FAR well below 5\%.  However, this comes at a cost --- the detection
probability is noticeably reduced as well.  The best choice for
$\log(P)$ depends on the application, and the exact form and quality
of $\log(P)$ also depends on the information that the application has
available.

The second method to improve the FAR exploits the grouping provided
by \PUMALINK.
If the input to \PUMALINK\ comprises more four observations of a
field, \PUMALINK\ will try to group quads that share detections and
are consistent with one another.  There are three possibilities for a
given quad: 1) detections are not shared with any other quad, 2)
detections are all shared by one of the groups that \PUMALINK\ forms
from proximate quads and the quad is absorbed into the group, and 3)
some detections are shared with another quad, but other
detections between these two quads are inconsistent with a single object.

In Case 1) the quad has to be considered on its own, as above.  In
Case 2) the quad has been absorbed into a group which has a much
higher probability of being real than a single quad --- in fact with
application of the ad-hoc probability test just described almost all
the groups of at least 5 detections that survive in the ATLAS
experiment are mis-linkages instead of bogus detections.

Case 3) normally indicates a mis-linkage.  A tracklet from one night
might link to more than one tracklet from the other epoch.  If
detections from the tracklets are inconsistent with being from the
same object because they are too far apart \PUMALINK\ will refuse to
merge these two quads or groups.  The user then needs to identify these
quads or groups that do share detections but are not linked together
and endeavor to decide which is correct.  Although both quads or
groups may have an acceptable aggregate probability and
\PUMA\ $\chi^2_\nu$, it may be possible to decide on one or the other by
which is more probable.  Another method is to start assigning detections
to objects using unambiguous quads or groups and by process of
elimination rule out cross-linkages because they contain a detection
already assigned to a different object.

As an example, there is a group of 5 detections in the
\PUMALINK\ between two of the ATLAS nights which looks like it is a
very good candidate to be real, visually and by $\log(P)$.  However,
this group of 5 shares a detection called {\tt 01a58995o0404o.12134}
with an even higher probability group of 7 but cannot merge with that
group because it also shares a detection called {\tt
  01a58999o0520o.6397} with a different group of 6.  The group of 7
and the group of 6 are more likely to be real than the group of 5 by
virtue of their size and having a higher probability.  (As it turns
out the group of 7 and the group of 6 are indeed two different kasts.)
If those two detections are removed from the pool by virtue of being
claimed by the group of 7 and group of 6, the group of 5 can be
eliminated as a mis-linkage.  A new, stand alone program that solves
this Sudoku-like puzzle of \PUMALINK\ mis-linkages could be very
powerful.

The third method for reducing false alarm is reclassifying.
Detections of real objects are usually distinguishable by eye from
detections of noise, even if they both have good $\log(P)$.  The ATLAS
classifier does not use a trained, machine learning test on all
detections but ATLAS does use such a test once MOPS links detections
as potential moving objects.  The performance is
excellent: 99.6\% TP for only 9\% FP \citep{2021PASP..133c4501C}.  The
reason to delay the application of a machine learning classifier until
\PUMALINK\ has been run is twofold: execution time is faster because
only a subset of detections are linked and need to be classified, and
as above with $\log(P)$ the aggregate classification of a group of
detections can be much more incisive than on each on individually.

Finally, validation of a potential linkage by finding other detections
that also belong essentially eliminates false alarms.  This can
be done on lower likelihood detections, such as the ATLAS 3-sigma
file by using \PUMA\ to interpolate all groups to the time and
observatory location of exposures and counting the number of nearby
detections.  In the case that there are more than two exposures per
epoch the detections may already exist in the input list, and helping
\PUMALINK's simple grouping function can provide a useful count.

Similarly, the experiments above only compare \PUMALINK\ performance on
2 nights at a time.  Bringing in extra nights so that
\PUMALINK\ is hierarchically combining 3 or 4 epochs of tracklets
reduces the false alarms and mis-linkages essentially to zero.  All of these
methods require knowledge that survey exposures exist at a given time
and sky location, so that the count of nearby detections can be
balanced by the count of exposures where the object was not seen.
The MPC isolated tracklet file, for example, does not lend itself to
statistical rejection of a possible tracklet, only confirmation.

All of these methods to suppress false alarms are application specific
and therefore cannot be part of the core \PUMALINK\ processing.  Even
in the case that many linkages are uncertain, perhaps because of few
detections over a large time span, the \PUMALINK\ output does offer a
very compact way of digesting all the detections into potential
linkages in a way that is easily saved and processed if and when new
data become available, or tested against old data or data from other
observatories.  \PUMALINK\ can accept detections from any time and any
observatory, the only requirement is that two detections must exist
per object at two different epochs.  (The \PUMA\ library itself
doesn't care at all about pairs or tracklets, of course.)

\section{Conclusion}

With four (soon five) sites the ATLAS project to find dangerous
asteroids is facing new challenges and opportunities for scheduling
and linking detections of asteroids.  In order to
succeed in this linking it is mandatory to have at least 4 detections,
at least for faint detections that do not have distinguishing
characteristics such as being trailed.  This requirement does not mean
that the observations have to come from the same observatory, nor that
the observations have to be closely contemporaneous.  The ATLAS system
can perform better when the different sites cooperate with one another
to cover the sky every night, so the system needs to be able to
efficiently link detections over time spans of one day or more, driven
by the various site's longitude and weather.

This paper describes the software that ATLAS uses to fit
orbits to sets of detections (\PUMA), and a new methodology to find
correct linkages between a set of undifferentiated observations
(\PUMALINK).

The \PUMA\ algorithm for fitting orbits to angular positions on the
sky is several orders of magnitude faster than the methods currently
in use by the community such as {\tt openorb} and {\tt find\_orb}, but
it does not sacrifice accuracy.  By taking advantage of the precision
of the angular positions and by avoiding the usual predictor-corrector
extrapolation strategy of differential equation integrators,
\PUMA\ can determine a 3D fit to 2D points on the sky in a
millisecond.  This opens the door for swift hypothesis testing on
different sets of detections that might or might not be of the same
object.

\PUMA\ uses carefully chosen polynomials to extrapolate beyond the
time interval of constraining observations, so the accuracy depends on
the ratio between the extrapolation interval $\Delta t$ and the
time span $\delta t$ of input observations.  For
$\Delta t/\delta t<1000$ \PUMA\ typically manages errors of 10's of
milliarcsec, with quadratic degradation as  $\Delta t/\delta\gg1000$.
Similarly, the ratio of the extrapolation interval and the ``collision
time'' $\rho/\dot\rho$ (the distance divided by the radial velocity)
governs the non-linearity of the trajectory.  \PUMA\ explicitly uses
the gravity of Sun, Earth, and Moon individually so it retains its
accuracy when $\dot\rho / \rho\, \Delta t \gtrsim 1$.

The new \PUMALINK\ algorithm operates on pairs of pairs of detections
(pairs of tracklets), deciding which are consistent with a real orbit.
\PUMALINK\ has similarities to other approaches, notably HelioLinC,
but it functions well at asteroid ranges of a small fraction of an AU.
\PUMALINK's effectiveness arises from the use of the Earth-Moon
barycenter as the origin for \PUMA, and it computes all possible
trajectories of a tracklet as a linear function of the unknown
distance and radial velocity.  The probability that some distance and
radial velocity can make two tracklets consistent at a given reference
time is then a simple, explicit formula from these linear function
coefficients.  It is therefore possible to test 10 million possible
tracklets against one another in a half hour of computer time.  The
accuracy of the \PUMA\ library ensures that the linkages are accurate,
even at distances much closer than the Moon, and the false alarm
rate is manageable, even for linkages of only 4 detections over
multiple days.

We demonstrated the performance of \PUMALINK\ on three datasets that
each span two weeks.  The first was a set of full ATLAS nights, with
two sites each observing a quarter sky in a declination band,
following one another every other day.  In addition to asteroids, these data
have a very large number of stationary variables and transients as
well as bogus detections.  The second dataset was an opposition field
(with many known asteroids) that various ATLAS sites observed at least
4 times every night with a quadrupled exposure time.  The third
dataset was an LSST simulation of all the moving objects that LSST
might see over 2 weeks.

For all three experiments we formed all pairs of nights, gave them to
\PUMALINK\ for linking, and then evaluated the performance.  We found
that \PUMALINK\ achieved a very high probability of detection (well
above 95\%) at a modest false alarm rate (less than 50\%).  The
detection probability is better when \PUMALINK\ is given a pair of
nights that have a small extrapolation ratio ($\Delta t/\delta
t\lesssim300$) but even for large extrapolation ratios ($\Delta
t/\delta t\gtrsim3000$) the detection probability is still above 95\%.
This false alarm rate refers to a minimal set of 4 detections to form
a quad between two nights.  Linkage groups with more than 4
detections have a much smaller false alarm rate, and if
\PUMALINK\ were given more than two nights of observation the false
alarm rate would be essentially zero.

Any linkage algorithm must balance detection probability, execution
time, and false alarm rate.  The optimal balance between these three
factors is application specific and therefore does not have a unique
solution to which \PUMALINK\ can aspire.  We discussed four methods to
augment the \PUMALINK\ output to trade these three against each other.
For example, we reduced the false alarm rate in the full night ATLAS
experiment by an order of magnitude at the same detection probability
by exploiting an ATLAS detection specific probability that all four
detections making a linkage are real.

For ATLAS our goal is to implement \PUMALINK, quadruple our normal
exposure times, and observe fields only 3 times every other night
instead of 4 times each night.  This should improve our detection
probability for nearby, approaching asteroids while also increasing
the discovery rate of ordinary NEOs.  The performance of such a
revised schedule and linking will be reported in a future paper.

\section{Acknowledgements}

Support for the ATLAS system was provided by NASA grants NN12AR55G,
80NSSC18K0284, 80NSSC18K1575, and this work was supported by NASA
grant 80NSSC23K0376.  We are grateful for the LSST data set provided
by Ari Heinze, the MPC ITF supplied by Matt Holman, and many
discussions with Larry Denneau.  This paper was substantially improved
following recommendations of an anonymous referee.



\clearpage

\begin{thebibliography}

\bibitem[Alard \& Lupton(1998)]{1998ApJ...503..325A} Alard, C. \& Lupton, R.~H.\ 1998, \apj, 503, 325. doi:10.1086/305984

\bibitem[Becker et al.(2015)]{hotpants} Becker, A., 2015, Astrophysics
  Source Code Library, http://adsabs.harvard.edu/abs/2015ascl.soft04004B

\bibitem[Bellm et al.(2019)]{2019PASP..131a8002B} Bellm, E.~C., Kulkarni, S.~R., Graham, M.~J., et al.\ 2019, \pasp, 131, 018002. doi:10.1088/1538-3873/aaecbe

\bibitem[Bernstein \& Khushalani(2000)]{2000AJ....120.3323B} Bernstein, G. \& Khushalani, B.\ 2000, \aj, 120, 3323. doi:10.1086/316868

\bibitem[\protect\citeauthoryear{Chambers et
    al.}{2016}]{2016arXiv161205560C} Chambers K.~C., Magnier, E.~A.,
  Metcalfe, N., et al., 2016, arXiv, arXiv:1612.05560 




\bibitem[Denneau et al.(2013)]{MOPS} Denneau, L., Jedicke, R., et
  al. 2013, \pasp, 125, 357

\bibitem[Drake et al.(2009)]{crts} Drake, A.~J., Djorgovski, S.~G., Mahabal, A., et al.\ 2009, \apj, 696, 870. doi:10.1088/0004-637X/696/1/870

\bibitem[Graham et al.(2019)]{2019PASP..131g8001G} Graham, M.~J., Kulkarni, S.~R., Bellm, E.~C., et al.\ 2019, \pasp, 131, 078001. doi:10.1088/1538-3873/ab006c

\bibitem[Granvik et al.(2009)]{2009M&PS...44.1853G} Granvik, M., Virtanen, J., Oszkiewicz, D., et al.\ 2009, \maps, 44, 1853. doi:10.1111/j.1945-5100.2009.tb01994.x

\bibitem[Heinze et al.(2018)]{2018AJ....156..241H} Heinze, A.~N., Tonry, J.~L., Denneau, L., et al.\ 2018, \aj, 156, 241. doi:10.3847/1538-3881/aae47f

\bibitem[Heinze et al.(2022)]{2022DPS....5450404H} Heinze, A., Eggl, S., Juric, M., et al.\ 2022, \dps


\bibitem[Holman et al.(2018)]{2018AJ....156..135H} Holman, M.~J., Payne, M.~J., Blankley, P., et al.\ 2018, \aj, 156, 135. doi:10.3847/1538-3881/aad69a

\bibitem[Ivezi{\'c} et al.(2019)]{2019ApJ...873..111I} Ivezi{\'c},
  {\v{Z}}., Kahn, S.~M., Tyson, J.~A., et al.\ 2019, \apj, 873,
  111. doi:10.3847/1538-4357/ab042c

\bibitem[Juric et al.(2017)]{2017ASPC..512..279J} Juric, M.,
  Kantor, J., Lim, K.-T., et al.\ 2017, Astronomical Data Analysis
  Software and Systems XXV, 512, 279. doi:10.48550/arXiv.1512.07914
  
\bibitem[Juric et al.(2021)]{2021DPS....5310106J} Juric, M., Eggl, S., Jones, L., et al.\ 2021, \dps

\bibitem[Kochanek et al.(2017)]{2017PASP..129j4502K} Kochanek, C.~S.,
  Shappee, B.~J., Stanek, K.~Z., et al.\ 2017, \pasp, 129,
  104502. doi:10.1088/1538-3873/aa80d9
  
\bibitem[Licandro et al.(2023)]{2023arXiv230207954L} Licandro, J., Tonry, J., Alarcon, M.~R., et al.\ 2023, arXiv:2302.07954. doi:10.48550/arXiv.2302.07954

\bibitem[Moeyens et al.(2021)]{2021AJ....162..143M} Moeyens, J., Juri{\'c}, M., Ford, J., et al.\ 2021, \aj, 162, 143. doi:10.3847/1538-3881/ac042b

\bibitem[NRC(2010)]{NRC2010} NRC 2010, Defending Planet Earth: Near-Earth-Object Surveys and Hazard Mitigation Strategies: Final Report. 2010. Washington, DC: The National Academies Press

\bibitem[Ofek et al.(2023)]{2023arXiv230404796O} Ofek, E.~O., Ben-Ami, S., Polishook, D., et al.\ 2023, arXiv:2304.04796. doi:10.48550/arXiv.2304.04796


\bibitem[Chyba Rabeendran \& Denneau(2021)]{2021PASP..133c4501C}
    Chyba Rabeendran, A. \& Denneau, L.\ 2021, \pasp, 133,
    034501. doi:10.1088/1538-3873/abc900
    
\bibitem[Tonry et al.(2018)]{ATLAS18} Tonry, J.~L., Denneau, L., Heinze, A.~N., et al.\ 2018, \pasp, 130, 064505 

\bibitem[Tonry et al.(2018)]{Refcat2} Tonry, J.~L., Denneau, L., Flewelling, H., et al.\ 2018, \apj, 867, 105. doi:10.3847/1538-4357/aae386

\bibitem[Zackay et al.(2016)]{zackaysub} Zackay, B., Ofek, E.~O., \& Gal-Yam, A.\ 2016, \apj, 830, 27. doi:10.3847/0004-637X/830/1/27


\end{thebibliography}
\end{document}